\documentclass[sigconf]{acmart}

\setcopyright{acmcopyright}
\copyrightyear{2023}
\acmYear{2023}
\acmDOI{XXXXXXX.XXXXXXX}

\acmConference[UIST '23]{2023 ACM Symposium on User Interface Software and Technology}{Oct 29--Nov 1,  2023}{California Bay Area, USA}
\acmPrice{15.00}
\acmISBN{978-1-4503-XXXX-X/18/06}

\copyrightyear{2023}
\acmYear{2023}
\setcopyright{rightsretained}
\acmConference[UIST '23]{The 36th Annual ACM Symposium on User Interface Software and Technology}{October 29-November 1, 2023}{San Francisco, CA, USA}
\acmBooktitle{The 36th Annual ACM Symposium on User Interface Software and Technology (UIST '23), October 29-November 1, 2023, San Francisco, CA, USA}
\acmDOI{10.1145/3586183.3606754}
\acmISBN{979-8-4007-0132-0/23/10}

\usepackage{enumitem,acmart-taps}
\usepackage{bbding}
\usepackage{float}
\usepackage{subcaption}
\usepackage{xcolor}
\usepackage{url}
\usepackage{multirow}
\usepackage{makecell}
\usepackage{bm}
\usepackage{amsthm}
\usepackage[vlined,linesnumbered]{algorithm2e}

\newtheorem{definition}{Definition}
\newtheorem{proposition}{Proposition}

\makeatletter
\newcommand{\removelatexerror}{\let\@latex@error\@gobble}
\makeatother

\renewcommand{\shortauthors}{Nair et al.}

\begin{document}

\title[Going Incognito in the Metaverse]{Going Incognito in the Metaverse: Achieving Theoretically Optimal Privacy-Usability Tradeoffs in VR}

\author{Vivek Nair}
\authornote{Equal contribution.}
\affiliation{%
  \institution{UC Berkeley}
  \city{Berkeley}
  \state{California}
  \country{USA}
}
\email{vcn@berkeley.edu}

\author{Gonzalo Munilla-Garrido}
\authornotemark[1]
\affiliation{%
  \institution{TU Munich}
  \city{Munich}
  \country{Germany}}
\email{gonzalo.munilla-garrido@tum.de}

\author{Dawn Song}
\affiliation{%
  \institution{UC Berkeley}
  \city{Berkeley}
  \state{California}
  \country{USA}}
\email{dawnsong@berkeley.edu}

\begin{abstract}
Virtual reality (VR) telepresence applications and the so-called ``metaverse'' promise to be the next major medium of human-computer interaction. However, with recent studies demonstrating the ease at which VR users can be profiled and deanonymized, metaverse platforms carry many of the privacy risks of the conventional internet (and more) while at present offering few of the defensive utilities that users are accustomed to having access to. To remedy this, we present the first known method of implementing an ``incognito mode'' for VR. Our technique leverages local $\bm{\varepsilon}$-differential privacy to quantifiably obscure sensitive user data attributes, with a focus on intelligently adding noise when and where it is needed most to maximize privacy while minimizing usability impact. Our system is capable of flexibly adapting to the unique needs of each VR application to further optimize this trade-off. We implement our solution as a universal Unity (C\#) plugin that we then evaluate using several popular VR applications. Upon faithfully replicating the most well-known VR privacy attack studies, we show a significant degradation of attacker capabilities when using our solution.
\end{abstract}

\ccsdesc[500]{Security and privacy}
\ccsdesc[500]{Security and privacy~Privacy protections}

\keywords{virtual reality, usable security, incognito mode, data harvesting, profiling, identification, private browsing, differential privacy}

\received{30 March 2023}
\received[accepted]{19 July 2023}

\maketitle

\section{Introduction}
\label{sec:Introduction}

Recent years have seen explosive growth in research and investment into the ``metaverse,'' which comprises immersive augmented and virtual reality (AR/VR) applications that claim to realize the next major iteration of the internet as a multi-user 3D virtual environment. Such platforms, by their very nature, transform every movement of their users into a stream of data to be rendered as a virtual character model for other users around the world.

Since at least the 1970s, researchers have understood that individuals exhibit distinct biomechanical motion patterns that can be used to identify them or infer their personal attributes \cite{cuttingrecognizing1977, kozlowskirecognizing1977}.
Thus, attention has rightly shifted toward the unique security and privacy threats that metaverse platforms may pose, with recent studies showing that seemingly-anonymous VR users can easily and accurately be profiled \cite{nair2022exploring} and deanonymized \cite{millerpersonal2020, nair2023unique} from just a few minutes of tracking data. They further show that while ``the potential scale and scope of this data collection far exceed what is feasible within traditional mobile and web applications'' \cite{nair2022exploring}, users are less broadly aware of security and privacy risks in VR than they are of similar risks in traditional platforms like social media \cite{noauthornational2022}.

Of course, data privacy challenges are not unique to VR. Nearly every major communications technology advancement of the past century has been accompanied by corresponding privacy risks.
For example, on the web, browser cookies pose a widely understood risk to privacy by attaching identifiers and tracking users across websites~\cite{cookies}. However, the maturation of web technologies has also brought an enhanced understanding of, and countermeasures to, such attacks, with technologies private browsing (or ``incognito'') mode in browsers providing users with vital defensive tools for reclaiming control of their data. By contrast, equivalent comprehensive privacy defenses have yet to be developed for the metaverse.
We thus find ourselves now in the dangerous situation of facing unprecedented privacy threats in VR while lacking the defensive resources we have become accustomed to on the web.

In this paper, we aim to begin addressing this disparity by designing and implementing the first ``incognito mode'' for VR. Our method leverages local $\varepsilon$-differential privacy to provide quantifiable resilience against known VR privacy attacks according to a user-adjustable privacy parameter $\varepsilon$.
In doing so, it allows for inherent privacy and usability \mbox{trade-offs} to be dynamically rebalanced, along a theoretically optimal continuum, according to the risks and requirements of each VR application, with a focus on the targeted addition of noise to those parameters which are most vulnerable.
We provide an open-source implementation of our solution as a Unity plugin, which we then use to replicate three existing VR privacy attack studies. Our results show a significant degradation of attacker capabilities when using our extension.

Finally, we provide statistical bounds for the perceived error that users may experience when using our technique. We argue that these bounds are well within the range that VR users can naturally adapt to according to past research on homuncular flexibility \cite{won2015homuncular}.

\eject

\noindent \textbf{Contributions}
\begin{enumerate}[leftmargin=*]
    \itemsep 0em
    \item We provide an $\varepsilon$-differential privacy framework for protecting a range of sensitive data attributes in VR (\S\ref{sec:Privacy_defenses}).
    \item We design and describe a concrete implementation of a modular ``VR Incognito Mode'' plugin for Unity (\S\ref{sec:VR_incognito}).
    \item We experimentally demonstrate the efficacy of our approach at defeating known VR privacy attacks (\S\ref{subsec:primary_secondary_attributes}).
\end{enumerate}
\vspace{-0.4em}

\section{Background \& Motivation}
\label{sec:background}

\vspace{-0.2em}

In this section, we aim to motivate the need for an ``incognito mode'' in VR. We begin by analyzing known privacy threats in VR, highlighting the ones we aim to address in this paper. Next, we present a comprehensive threat model to illustrate which threats are feasibly mitigated by software-based client-side defenses. We then briefly discuss private web browsing to draw an analogue to the goals of this paper. Finally, we introduce differential privacy and randomized response, the theoretical building blocks for our solution and proof-of-concept implementation.

\vspace{-0.5em}

\subsection{VR Privacy Attacks}
\label{subsec:vr_privacy_attacks}

\vspace{-0.2em}
Our primary motivation for pursuing this research is the breadth of prior work demonstrating compelling privacy risks within the metaverse. Among the extant research in this domain are papers ranging from high-level literature reviews on VR privacy~\cite{convergeneceVRsocialmedia, privimplicationsVR, privmetaverse, metaverseprivacy1, metaverseintro, VRdisclosureinfo, VRreviewanony, dickbalancing2021, deguzmansecurity2020, garrido2023sok} to targeted risk assessment frameworks \cite{VRattackeducation, VRattackeducation2} and privacy guidelines \cite{VRprivtutorial}. Focusing specifically on the technical works, we note the following relevant studies: \medskip

\noindent \textbf{Eye and Body Tracking.} Several works focus specifically on the security and privacy of eye tracking \cite{10.1145/3313831.3376840, krogerwhat2020}. We place a limited emphasis on eye tracking in this paper, as features such as foveated rendering are not yet widespread, and there are already known effective countermeasures \cite{VReyetrackpriv, VReyetrackpriv2, VReyetrackpriv3, snowpixeleyetracking, 263891}. We similarly set aside the privacy of full-body motion capture systems \cite{VRprivbody1, VRprivbody2, VRprivbody3, VRprivbody4}. Instead, we focus on the simple setup of a headset plus two handheld controllers, as is found on most consumer VR devices today. \medskip

\noindent \textbf{Comprehensive Attacks.} The attacks most relevant to this paper are those of the 2020 Miller et al. ``TTI''\footnote{The study was conducted at The Tech Interactive (TTI) museum in San Jose.} study \cite{millerpersonal2020}, the 2022 Nair et al. ``MetaData'' study \cite{nair2022exploring}, and the 2023 Nair et al. ``50k'' study \cite{nair2023unique}. First, the TTI study demonstrated that 511 seemingly-anonymous VR users could be deanonymized with 95\% accuracy from just 5 minutes of tracking data. The MetaData study expanded on this result, showing that a malicious VR application can also ascertain more than $25$ private data points from its users, including various environmental, demographic, and anthropometric attributes. Finally, the 50k study showed that 55,541 VR users can be uniquely identified with 94.33\% accuracy from 100 seconds of motion data collected from the popular ``Beat Saber'' VR game. Together, the below attributes are those that the literature suggests can be harvested from VR users and that our techniques aim to protect:

\begin{itemize}[leftmargin=*]
    \itemsep 0em
    \item \textbf{Anthropometrics:} height, wingspan, arm lengths, fitness, interpupillary distance, handedness, reaction time~\cite{nair2022exploring}.
    \item \textbf{Environment:} room size, geolocation~\cite{nair2022exploring}.
    \item \textbf{Technical:} tracking/refresh rate, device model~\cite{nair2022exploring, 277092}.
    \item \textbf{Demographics:} gender, age, ethnicity, income~\cite{nair2022exploring}.
    \item \textbf{Identity}~\cite{millerpersonal2020, 9756791, nair2023unique}.
\end{itemize}

\subsection{Metaverse Threat Model}
\label{subsec:threat_model}


We present a threat model to contextualize our contributions within the broader ecosystem of VR privacy. Our model is adapted from the standard model proposed by Garrido et al. \cite{garrido2023sok}. We consider a \textit{target user} who interacts with the metaverse over multiple \textit{usage sessions}. The parties which could plausibly observe a session are: \medskip

\begin{itemize}[leftmargin=*]
\item A \textbf{(I) Hardware Attacker}, which controls the hardware and firmware of the target user's VR device, and thus has access to raw sensor data from the VR hardware.
\item A \textbf{(II) Client Attacker}, which controls the client-side VR application running on the target user's device, and thus has access to data provided by the device APIs.
\item A \textbf{(III) Server Attacker}, which controls the external server used to facilitate multi-player functionality, and thus receives a stream of telemetry data from the client.
\item A \textbf{(IV) User Attacker}, which represents another end-user of the same VR application, and thus naturally receives from the server a stream of data about the target user. \medskip
\end{itemize}

\begin{figure}[h]
\includegraphics[width=0.75 \linewidth]{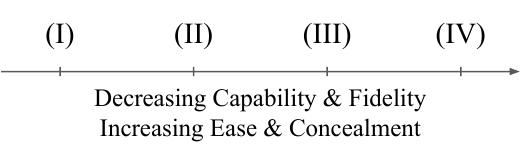}
\centering
\vspace{-1.5em}
\caption{Continuum of VR privacy attackers.}
\label{fig:spectrum}
\end{figure}

In our model, the goals of an attacker are to correctly observe attributes of the target user, or to identify them across multiple sessions. Fig. \ref{fig:spectrum} shows that the four attackers lie on a continuum; the later attackers have less privilege and attack accuracy, but can more easily conceal their attacks. Generally, each attacker inherits a subset of the capabilities of the previous attackers as data streams become increasingly processed and filtered at each step.

\vspace{-0.5em}
\vspace{-0.5em}

In this paper, we present algorithmic statistical defenses for the vulnerable attributes of \S\ref{subsec:vr_privacy_attacks} that can be implemented at either the device firmware or client software level. Tab. \ref{tab:coverage} shows the attackers covered by each implementation possibility.
In practice, lacking any special access to VR device firmware, our evaluated systems were all implemented at the software level.

\vspace{-0.5em}

\aptLtoX[graphic=no,type=html]{ \begin{table}[H]
\centering
\begin{tabular}{lcccc}
\toprule
 & \multicolumn{4}{c}{\textbf{Attackers}} \\ 
\colrule
 & \multicolumn{1}{c|}{\textbf{I}} & \multicolumn{1}{c|}{\textbf{II}} & \multicolumn{1}{c|}{\textbf{III}} & \multicolumn{1}{c|}{\textbf{IV}} \\ \hline
\multicolumn{1}{|l|}{\textit{Software Incognito}} & \multicolumn{1}{c|}{} &\multicolumn{1}{c|}{}  & \multicolumn{1}{c|}{\Checkmark} & \multicolumn{1}{c|}{\Checkmark} \\ \hline
\multicolumn{1}{|l|}{\textit{Firmware Incognito}} &  \multicolumn{1}{c|}{} & \multicolumn{1}{c|}{\Checkmark} & \multicolumn{1}{c|}{\Checkmark} & \multicolumn{1}{c|}{\Checkmark} \\ 
\bottomrule
\end{tabular}%
\caption{Coverage of proposed defenses.}
\label{tab:coverage}
\end{table} }{ \begin{table}[H]
\centering
\begin{tabular}{lcccc}
 & \multicolumn{4}{c}{\textbf{Attackers}} \\ \cline{2-5} 
 & \multicolumn{1}{c|}{\textbf{I}} & \multicolumn{1}{c|}{\textbf{II}} & \multicolumn{1}{c|}{\textbf{III}} & \multicolumn{1}{c|}{\textbf{IV}} \\ \hline
\multicolumn{1}{|l|}{\textit{\begin{tabular}[c]{@{}l@{}}Software Incognito\end{tabular}}} & \multicolumn{1}{c|}{} & \multicolumn{1}{c|}{} & \multicolumn{1}{c|}{\Checkmark} & \multicolumn{1}{c|}{\Checkmark} \\ \hline
\multicolumn{1}{|l|}{\textit{\begin{tabular}[c]{@{}l@{}}Firmware Incognito\end{tabular}}} & \multicolumn{1}{c|}{} & \multicolumn{1}{c|}{\Checkmark} & \multicolumn{1}{c|}{\Checkmark} & \multicolumn{1}{c|}{\Checkmark} \\ \hline
\end{tabular}%
\caption{Coverage of proposed defenses.}
\label{tab:coverage}
\end{table} }

\vspace{-1.5em}

Overall, the ``VR incognito mode'' defenses proposed in this paper are unable to address the threat of hardware and firmware level attackers. We argue that this is a necessary concession of a software-based defense, and that unlike the client, server, and user attackers we cover, hardware and firmware attacks can be discovered via reverse engineering.
Still, in an ideal world, VR devices would contain hardware-based mechanisms for ensuring user privacy. As it stands, VR firmware is tightly controlled and not alterable by researchers without cooperation from OEMs, who are presently disincentivized from implementing hardware-level privacy protections.

\subsection{Private Web Browsing}
\label{subsec:private_web_browsing}

We now detour briefly to the more mature field of private web browsing to seek inspiration from the web privacy solutions which have stood the test of time.

The research community has surveyed the field of web privacy \cite{webbrowsingprotections, tsalisexploring2017}, and identified observable attributes ranging from tracking cookies \cite{cookies} and HTTP headers \cite{fingerprintingsurvey} to browsing histories \cite{privsniffattackbrowser} and motion sensor data \cite{motionprivwebattack}. As in VR, these attributes can be combined to achieve profiling \cite{englehardtcookies2015, profiling}, fingerprinting \cite{fingerprintingsurvey} and deanonymization \cite{281290}.
Further, the attack model used by web privacy researchers resembles the metaverse threat model presented in \S\ref{subsec:threat_model}, with most defenses focusing on web servers and other users, some on client-side applications, and relatively few on the underlying hardware.

In response to these threats, proposed solutions have included proxies, VPNs~\cite{kaaniche2020privacy}, Tor~\cite{torproject, deanontor}, and, of course, private browsing or ``incognito'' mode in browsers, as well as dedicated private browsers and search engines, e.g., Brave~\cite{brave} and DuckDuckGo~\cite{DuckDuckGo}. Of these solutions, ``incognito mode'' stands out due to its ease of use: a wide range of defensive modifications to protocols, APIs, cookies, and browsing history can all be deployed with a single click \cite{10.5555/1929820.1929828}. Due perhaps to this outward simplicity, surveys of web privacy protections used in practice have found private browsing mode to be by far the most popular at 73\% adoption \cite{219416}.

In summary, web privacy is highly analogous to metaverse privacy; although the data attributes being protected are vastly different, the threat of combining attributes to profile and deanonymize users is a constant, as is the threat model used to characterize both fields. On the other hand, the size and scope of data collection in VR potentially exceed that of the web \cite{nair2022exploring}, while users are simultaneously less aware of the threat in VR \cite{10.1145/3385956.3418967}, and the equivalent privacy tools are not generally available. We are motivated by the popularity of incognito mode on the web to seek an equivalent for VR, with the same fundamental goal as in browsers: allowing users, at the flick of a switch, to become harder to trace across sessions.

\subsection{Differential Privacy}
\label{subsec:differential_privacy}

Having established our motivation for pursuing a metaverse equivalent to ``incognito mode,'' we now lay out the tools necessary to enable its realization. Chief among these is differential privacy~\cite{DPoriginal}, which provides a context-agnostic mathematical definition of privacy that statistically bounds the information gained by a hypothetical adversary from the output of a given function $\mathcal{M}(\cdot)$:

\begin{definition}
\label{def:DP}
  ($\varepsilon$-Differential Privacy~\cite{dworkalgorithmic2013}). A randomized function $\mathcal{M}(\cdot)$ is $\varepsilon$-differentially private if for all input datasets $D$ and $D'$ differing on at most one element, and for all possible outputs $\mathcal{S} \subseteq
  \mathit{Range}(\mathcal{M})$:
      	$\mathrm{Pr}[\mathcal{M} (D) \in \mathcal{S}] \leq e^\varepsilon \times \mathrm{Pr}[\mathcal{M} (D') \in \mathcal{S}]$.
\end{definition}

A function $\mathcal{M}(\cdot)$ fulfills differential privacy if its outputs with and without the presence of an individual input element are indistinguishable with respect to the privacy parameter $\varepsilon \geq 0$.
In practice, a randomized function $\mathcal{M}(\cdot)$ typically ensures differential privacy by adding calibrated random noise to the output of a deterministic function, $\mathcal{M}(x) = f(x) + \mathrm{Noise}$.
Lower $\varepsilon$ values correspond to higher noise, making it harder to distinguish outputs and strengthening the privacy protection.
In addition to $\varepsilon$, the required noise is affected by the \emph{sensitivity} ($\Delta$) of the deterministic function. 

Another aspect worth highlighting is \emph{sequential composition}~\cite{dworkalgorithmic2013}: if $\mathcal{M}(\cdot)$ is computed $n$ times over $D$ with $\varepsilon_i$, the total \textit{privacy budget} consumed is $\sum \varepsilon_i$.
Thus, users' attributes become less protected with every query execution.
Differentially private outputs are also \emph{immune to post-processing}~\cite{dworkalgorithmic2013}; an adversary can compute any function on the output (e.g., rounding) without reducing privacy.

In practice, differential privacy can be used \emph{centrally}, whereby a server adds noise to an aggregation function computed over data from multiple clients, or \emph{locally}, whereby clients add noise to data points before sharing them with a server. While local differential privacy is noisier than the central variant, it also requires less trust of the server. Since servers are considered potential adversaries in our threat model (\S\ref{subsec:threat_model}), we use local differential privacy to protect VR users in this paper. Specifically, we implement local differential privacy using the Bounded Laplace Mechanism~\cite{boundedtruncatedIBM, dworkalgorithmic2013} for continuous attributes and randomized response~\cite{warner1965randomized} for Boolean attributes. \newline

\vspace{-0.6em}

\noindent \mbox{\textbf{Bounded Laplace Mechanism.}} The Laplace mechanism~\cite{dworkalgorithmic2013}, also known as the ``workhorse of differential privacy,''~\cite{boundedtruncatedIBM} is a popular method of implementing local differential privacy for continuous attributes.
Laplacian noise satisfies a stronger notion of  $\varepsilon$-differential privacy than Gaussian noise, which only satisfies a weaker ($\varepsilon$, $\delta$)-differential privacy {\cite{zhao2019reviewing}}.
However, its unbounded noise can yield semantically absurd edge cases (e.g., a negative value for the height attribute).
Thus, in this paper, we use the Bounded Laplace mechanism~\cite{boundedtruncatedIBM}, which transforms the noise distribution according to the privacy parameters and deterministic value, then samples outputs until a value falls within pre-determined bounds without compromising differential privacy.
Inputs that fall outside the bounds are automatically clamped to the nearest bound.
Additionally, we employ the modified sampling technique of Holohan~et.~al~\cite{randomsamplingDP} to avoid a known vulnerability associated with the use of finite floating-point in other differential privacy implementations~\cite{mironovsignificance2012}. 
\newline

\vspace{-0.6em}

\noindent \textbf{Randomized Response.} To achieve local differential privacy for Boolean attributes, we can apply the randomized response method from Warner~\cite{warner1965randomized}: (i) the client flips a coin, (ii) if heads, the client sends a truthful response, (iii) else, the client flips a coin again and sends ``true'' if heads and ``false'' if tails. This method has been shown to be (${\varepsilon=\ln3}$)-differentially private with a fair coin~\cite{dworkalgorithmic2013}, though one can vary $\varepsilon$ by changing the bias of the first coin flip. 

\vspace{-0.25em}

\subsection{Homuncular Flexibility}
\label{subsec:homuncular_flexibility}

While differential privacy can be used to quantifiably address the problem of data leakage from VR telemetry, it does so by introducing noise to the VR data, thus potentially degrading the user experience.
However, past research on ``homuncular flexibility'' has shown  that users can learn to control bodies that are different from their own, particularly in virtual reality \cite{won2015homuncular, abtahi2022beyond}.
Thus, the remainder of this work focuses on deploying differential privacy in VR in a way that users can rapidly learn to ignore.
By transforming the virtual object hierarchy according to known usable non-linear interaction techniques (e.g., the Go-Go technique \cite{poupyrev1996go}), the corresponding attributes (e.g., wingspan) can be obscured while allowing users to flexibly adapt to their new environment.
\section{VR Privacy Defenses}
\label{sec:Privacy_defenses}

In this section, we provide a differentially-private framework for user data attribute protection in VR. We define each attribute defense in terms of abstract coordinate transformations, without regard to any specific method of implementation.
Later, in \S\ref{sec:VR_incognito}, we describe a concrete system for implementing these defenses within VR applications via a universal Unity plugin.

Our ``incognito mode'' defenses aim to prevent adversaries from tracking VR users across sessions in the metaverse.
In practice, this means limiting the number of data attributes adversaries can reliably harvest from users and use to infer their identity.
Local differential privacy (LDP) is the primary tool that allows us to achieve this with a mathematically quantifiable degree of privacy.
LDP has the effect of significantly widening the range of attribute values observed by an adversary given a particular ground truth attribute value of a user.
In doing so, it ensures that the observable attribute profile of a user always significantly overlaps with that of at least several other users, thus making a precise determination of identity infeasible.
The noise added by LDP may have some negative impacts on user experience, as is the case with incognito mode in browsers.
However, users can tune the privacy parameter ($\varepsilon$) to reduce the impact of noise on user experience as required.

Upon initiating a new metaverse session (i.e., connecting to a VR server), the defenses generate a random set of ``offset'' values, which are then used throughout the session to obfuscate attributes within the VR telemetry data stream through a set of deterministic coordinate transformations.
The re-randomization of offset values at the start of each session ensures that all usage sessions of a user are statistically unlinkable.\footnote{Methods for tracking users that are not unique to VR (such as via their IP addresses) are not considered to be within the scope of this paper; defenses like VPNs are widespread.}
On the other hand, these offsets remain consistent within a session to ensure adversaries never receive more than one view of sensitive attribute values.

\begin{figure}[h]
\includegraphics[width=0.25 \linewidth]{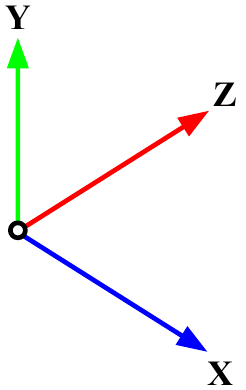}
\centering
\caption{Left-handed, Y-up Unity 3D coordinate system.}
\label{fig:coordinates}
\end{figure}

What follows are the specific differentially-private coordinate transformations that protect user data attributes (and thus allow them to ``go incognito'') in VR.
While for simplicity this section considers the protections for each attribute in isolation, in practice, our implementation uses a relative transformation hierarchy to allow any set of enabled defenses to seamlessly combine with each other (see \S\ref{subsec:defense_implementation}).
The coordinates used throughout this paper refer to the left-handed, Y-up Unity coordinate system, pictured in Fig. \ref{fig:coordinates}. 

\subsection{Preliminaries}

In our setting, LDP protects against adversaries with knowledge of observed attributes across all user sessions except for the current session of a target user ($D'$).
Sequential composition allows us to provide an upper bound for a user's privacy budget as the sum of each $\varepsilon$ used per attribute.

We identified the Bounded Laplace mechanism~\cite{boundedtruncatedIBM} as our tool of choice for protecting continuous attributes like \emph{height}, \emph{wingspan}, and \emph{room size} in VR because it produces random noise centered around the sensitive value (e.g., \emph{height}) while preserving the semantic consistency of the attribute (e.g., \emph{height} $> 0$).
The Laplacian noise distribution is preferable over, e.g., simply imbuing uniformly distributed random noise, because it has the property of minimizing the mean-squared error of any attribute at a given privacy level ($\varepsilon$) \cite{ldpmin}, thereby minimizing its impact the user experience.

Where Boolean attributes are concerned, we use randomized response \cite{warner1965randomized} with a weighted coin to provide $\varepsilon$-differential privacy for chosen values of $\varepsilon$.
The use of randomized response over simpler mechanisms (e.g., a single coin flip)  aligns Boolean attributes with the same $\varepsilon$-differential privacy framework as continuous attributes, and thus allows the $\varepsilon$ values of multiple attributes to be combined into a single ``privacy budget'' if desired.

\medskip

\noindent Throughout this paper, we use the following standard \mbox{variable} notation in our algorithm statements:

\smallskip

\begin{itemize}[leftmargin=*]
    \item $v$: sensitive deterministic value (``ground truth'')
    \item $(l_v, u_v)$: population bounds of $v$
    \item $\varepsilon \geq 0$: differential privacy parameter
    \item $p$: randomized response coin bias
    \item $(x_h, y_h, z_h)$: headset coordinates
    \item $(x_r, y_r, z_r)$: right controller coordinates
    \item $(x_l, y_l, z_l)$: left controller coordinates
\end{itemize}

\medskip

For a given attribute $a$ (e.g., $\mathit{height}$), we use $a'$ (e.g., $\mathit{height}'$) to denote the LDP-protected value an adversary observes. Our use of local differential privacy requires $\Delta$ to cover the entire range of the bounded interval $[l, u]$ ($\Delta = |u-l|$). Alg. \ref{alg:defenses} contains helper functions for the mechanisms discussed here that will be used throughout \S\ref{sec:Privacy_defenses}.

\medskip


\aptLtoX[graphic=no,type=html]{ \includegraphics{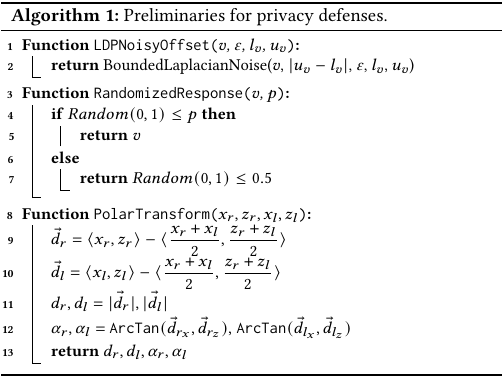} }{ \SetKwComment{Comment}{// }{}
\RestyleAlgo{ruled}
\begin{algorithm}[!h]
\small
\DontPrintSemicolon
\caption{Preliminaries for privacy defenses.}
\label{alg:defenses}



\SetKwFunction{Fsub}{LDPNoisyOffset}
\SetKwProg{Fn}{Function}{:}{}
\Fn{\Fsub{$v$, $\varepsilon$, $l_v$, $u_v$}}{
    \KwRet BoundedLaplacianNoise($v$, $|u_v - l_v|$, $\varepsilon$, $l_v$, $u_v$)\;
}

\SetKwFunction{Fsub}{RandomizedResponse}
\SetKwProg{Fn}{Function}{:}{}
\Fn{\Fsub{$v$, $p$}}{
    \uIf{$Random(0,1) \leq p$}{ 
      \KwRet $v$
    }
    \Else{\KwRet $Random(0,1) \leq 0.5$}
}

\SetKwFunction{Fsub}{PolarTransform}
\SetKwProg{Fn}{Function}{:}{}
\Fn{\Fsub{$x_r, z_r, x_l, z_l$}}{
    
    $\vec{d}_r  = \langle x_r, z_r \rangle - \langle \dfrac{x_r+x_l}{2}, \dfrac{z_r+z_l}{2} \rangle$
    
    $\vec{d}_l = \langle x_l, z_l \rangle - \langle \dfrac{x_r+x_l}{2}, \dfrac{z_r+z_l}{2} \rangle$
    
    $d_r, d_l = |\vec{d}_r|, |\vec{d}_l|$
    
    $\alpha_r, \alpha_l = $ {\tt ArcTan}$(\vec{d}_{r_x}, \vec{d}_{r_z}),$ {\tt ArcTan}$(\vec{d}_{l_x}, \vec{d}_{l_z})$
    
    \KwRet  $d_r, d_l, \alpha_r, \alpha_l$
}
\end{algorithm}

\vspace{-1.5em} }

\subsection{Continuous Attributes}
\label{subsec:continuous_Anthropocentrics}

Using the preliminaries established above, and in particular the Bounded Laplace mechanism, we now describe coordinate transformations for protecting continuous attributes in VR.
Each defense begins by calculating an $\mathit{offset}$ using the {\tt LDPNoisyOffset} helper function before diverging into two distinct categories: \emph{additive offset} defenses, which protect attributes such as interpupillary distance (IPD) that are not expected to change over the course of a session, and \emph{multiplicative offset} defenses, which protect attributes like observed height that might be updated each frame.

\medskip

\bigskip

\noindent \textbf{Additive Offset}

\smallskip

There are two continuous attributes that we can protect by simply adding a fixed $\mathit{offset}$ value to the ground truth as a one-time transformation: \emph{interpupillary distance} (IPD), and \emph{voice pitch}. The use of an additive offset is sufficient to protect these attributes without impacting usability due to the relatively static nature of such attributes throughout a session, with the resulting defenses being shown in Alg. \ref{alg:defenses_continuous_fixed}. \newline

\medskip

\noindent{\textit{IPD}.} 
We start with IPD as it is amongst the easiest attributes to defend due to the fact that it should not reasonably be expected to change during a session.
Our suggested countermeasure to attacks on IPD defends the player by scaling their avatar such that when an adversary measures the gap between their left and right eyes, the distance will correspond to a differentially private value. \newline

\medskip

\noindent{\textit{Voice Pitch}.} An attacker can also fingerprint a VR user by observing the median frequency of their speech as measured by a microphone on their VR device, which they can use in particular to infer a user's gender in addition to simply being a unique identifier.
Thus, we suggest pitch-correcting the voice stream according to the differentially-private $\mathit{offset}$.
As with IPD, the attacker can now only observe a differentially private $\mathit{pitch} + \mathit{offset}$ value.
Incidentally, we found that this defense is also sufficient to confuse machine learning models which attempt to infer the user's ethnicity based on their accent (see \S\ref{sec:results}), though that effect may be less resilient.

Studies which focus entirely on speech privacy \cite{speechprivacyieee} have presented more sophisticated techniques for obfuscating voice than the ones discussed here, but we include this differentially-private defense for completeness given the inclusion of speech attributes in VR attack papers \cite{nair2022exploring}. \newline

\medskip


\aptLtoX[graphic=no,type=html]{ \includegraphics{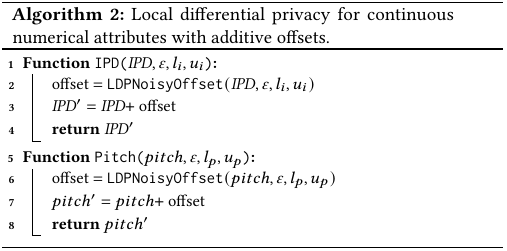} }{ \SetKwComment{Comment}{// }{}
\RestyleAlgo{ruled}
\removelatexerror
\begin{algorithm}[h]
\small
\DontPrintSemicolon
\caption{Local differential privacy for continuous numerical attributes with additive offsets.}
\label{alg:defenses_continuous_fixed}


\SetKwFunction{Fsub}{IPD}
\SetKwProg{Fn}{Function}{:}{}
\Fn{\Fsub{$\mathit{IPD}, \varepsilon, l_{i}, u_{i}$}}{
    offset $=$ {\tt LDPNoisyOffset}$(\mathit{IPD}, \varepsilon, l_{i}, u_{i})$\;
    $\mathit{IPD}' = \mathit{IPD} +$ offset\;
    \KwRet $\mathit{IPD}'$
}

\SetKwFunction{Fsub}{Pitch}
\SetKwProg{Fn}{Function}{:}{}
\Fn{\Fsub{$pitch, \varepsilon, l_{p}, u_{p}$}}{
    offset $=$ {\tt LDPNoisyOffset}$(pitch, \varepsilon, l_{p}, u_{p})$\;
    $pitch' = pitch +$ offset\;
    \KwRet  $pitch'$
}
\end{algorithm} }


\noindent{\textbf{Multiplicative Offset}}

\begin{figure}[h]
\includegraphics[width=\linewidth]{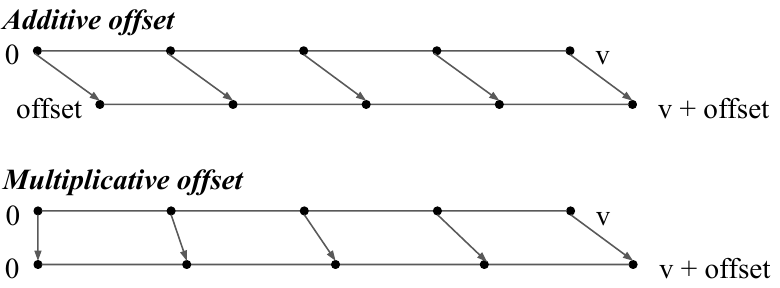}
\centering
\caption{Additive vs. multiplicative offset transformations.}
\label{fig:offsets}
\end{figure}

We now turn our attention to the bulk of attributes for which a multiplicative offset is required. Consider, for example, the case of \emph{wingspan}, where the perceived distance between a user's hands should appear to be $0$ when their hands are touching, but should reflect $\mathit{wingspan} + \mathit{offset}$ when their hands are fully extended. Simply adding $\mathit{offset}$ to the distance in all cases, as per the \emph{additive offset} approach, is insufficient to achieve this property. Instead, we scale the entire range of values by $v'/v$ as shown in Fig. \ref{fig:offsets}. As a result, observable attributes attain a differentially-private value at their extremes, while their zero-point is maintained. We present in this section \emph{multiplicative offset} defenses for a variety of attributes, as summarized in Alg. \ref{alg:defenses_continuous_dynamic}.

\vspace{-0.2em}

\vspace{-0.1em}

\noindent{\textit{Height}.} 
A typical method for inferring the height of a VR user is to record the y-coordinate of the VR headset ($y_h$) over the course of a session, and then use the highest observed coordinate (or, e.g., the $99$th percentile) as a direct linear correlate of height.
This attack is effective because $y_h = \mathit{height}$ when a user is standing upright, which they generally are for a large portion of their session.

While one may be tempted to simply adjust $y_h$ by $\mathit{offset}$ at all times, doing so could cause the relative error of a fixed offset can grow to become disproportionate in applications where users are required to get close to the ground. In fact, in an extreme scenario where a user decides to lie flat on the ground, an adversary may observe $y'_h = 0 + \mathit{offset}$, which could defeat the privacy of this method by revealing $\mathit{offset}$.

\begin{figure}[h]
\includegraphics[width=\linewidth]{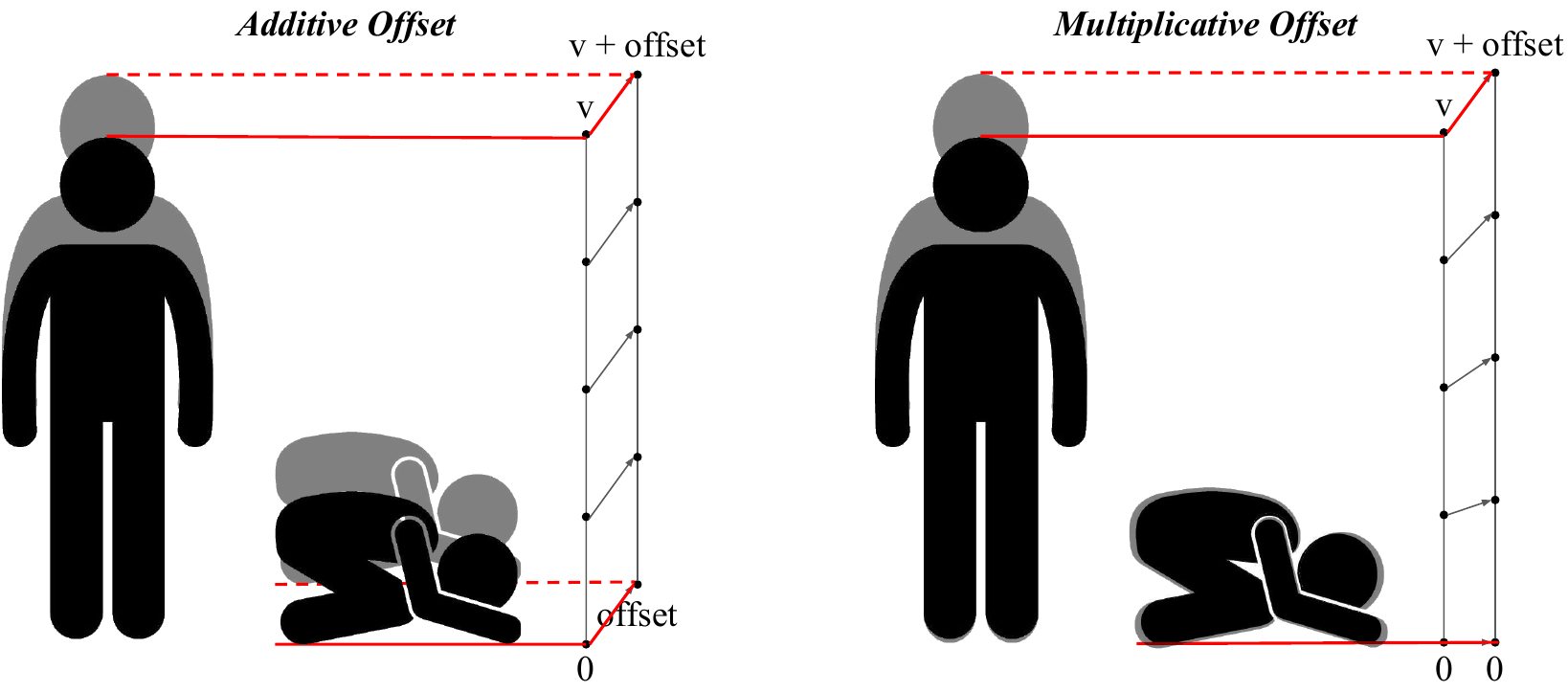}
\centering
\caption{Use of additive vs. multiplicative offset for height.}
\label{fig:height}
\end{figure}

Therefore, our suggested countermeasure is to use a \emph{multiplicative offset}, whereby $y'_h=y_h*(height'/height)$. When $y_h=height$, the adversary now observes the differentially-private value $y'_h = height + \mathit{offset}$, while $y'_h=0$ when $y_h=0$ as shown in Fig. \ref{fig:height}. We also suggest adjusting $y_r$ and $y_l$ such that the relative distance between the user's head and hands appears to remain unchanged.

\vspace{-0.2em}

\noindent{\textit{Squat Depth}.} 
Prior works have shown that an adversary can assess a proxy of a user's physical fitness by covertly prompting the users to squat and measuring their \emph{squat depth}, i.e., $depth=height - y_h$, where $y_h$ is the lowest headset coordinate recorded during the squat.
The aim of this defense is to ensure that an adversary can only observe a differentially private $depth$ value.
While this could be achieved by setting a strict lower bound on $y_h$, doing so has the potential to be disorienting and could potentially have a negative impact on the VR user experience perspective. Instead, our suggested defense offsets $y_h$ using the following transformation (independent of any defenses to $\mathit{height}$):

\bigskip

\begin{center}
$y'_h = height - (height - y_h)*(depth'/depth)$ \\
\end{center}

\bigskip

Consequently, that $y'_h$ smoothly transitions from $\mathit{height}$ to\break
${\mathit{height} - \mathit{depth} + \mathit{noise}}$ as $y_h$ goes from $\mathit{height}$ to ${\mathit{height} - \mathit{depth}}$,\break obscuring the user's actual squat depth.

\bigskip

\noindent{\textit{Wingspan}.}
The wingspan attribute is harvested in a similar way to height, with an adversary monitoring the distance $d$ between the left and right controllers over the course of a usage session and using the maximum observed value of $d$ as a strong correlate of the user's wingspan.
A VR application could require a user to fully extend their arms for seemingly legitimate gaming purposes, thus revealing their wingspan to potential attackers. The defense must therefore modify the observed distance $d$ when the user's arms are extended. However, as discussed at the start of this section, simply adding a fixed offset to $d$ does not allow $d=0$ when the user's hands are touching, which is desirable for UX.

In function {\tt Wingspan} of Alg.~\ref{alg:defenses_continuous_dynamic}, we formally introduce our recommended defense, where $\mathit{arm}_R$ and $\mathit{arm}_L$ are the arm length measurements in VR.
As with our protection of squat depth, we ensure that the noise scales smoothly to preserve the user experience.
As a result, when the user's hands are at the same coordinates, the observed distance is $0$; thus, when the user touches their physical hands, the virtual hands also touch.
On the other hand, when the arms are extended completely, the real-time distances between the controllers and their midpoint become $d_r = \mathit{arm}_R$ and $d_l= \mathit{arm}_L$, where $d_r + d_l = span$.
In such a position, the observed wingspan becomes differentially private:

\bigskip

\noindent \centerline{$\mathrm{offset} = \dfrac{d_r}{\mathit{arm_R}}* \dfrac{\mathit{span}'}{2} - d_r + \dfrac{d_l}{\mathit{arm_L}}* \dfrac{\mathit{span}'}{2} - d_l$}
\noindent \centerline{$\therefore \tfrac{span'}{2} - d_r + \tfrac{span'}{2} - d_l = span' - (d_r + d_l) = \mathrm{offset}$}

\bigskip

The defense adds half the total offset to each arm.
Consequently, the adversary will only observe a differentially private wingspan value when using the controllers' coordinates (($x_r, z_r$) and ($x_l, z_l$)) to calculate the distance:

\bigskip

\noindent \centerline{$|\langle x_r, z_r \rangle - \langle x_l, z_l \rangle| = \tfrac{span'}{2} + \tfrac{span'}{2} = span'$}

\bigskip

In VR research, this is known as the ``go-go technique'' \cite{poupyrev1996go}; here, we use a small scale factor to obscure the user's wingspan (rather than to extend reach). As with the other multiplicative offset defenses, post-processing immunity protects the sensitive values when multiplied by $\tfrac{w}{v} \in [0,1]$, and the adversary can only learn $span'$ from the observed distances in the range $[0, span']$.

\noindent{\textit{Arm Length Ratio}.}
If an adversary manages to measure the wingspan of a user, determining the arm length ratio is possible by using the headset as an approximate midpoint.
As function {\tt Arms} of Algorithm~\ref{alg:defenses_continuous_dynamic} shows, the corresponding defense is almost equivalent to that of the user's \emph{wingspan}, but while the wingspan protection adds noise symmetrically to both arms, in this case, we add noise asymmetrically to obfuscate the ratio of arm lengths. This reflects a unique deployment of the go-go technique with different scale factors used for each arm to obscure length asymmetries.

\bigskip


 
\noindent{\textit{Room Size}.}
Lastly, previous works have demonstrated that an adversary can determine the dimensions of a user's play area by observing the range of their movement. Once again, an additive offset would fail to defend against this attack by simply shifting the user's position rather than affecting their movement range.
We therefore employ a similar technique as with the other multiplicative offset transformations in that the dynamic noise at the center of the room is $0$, which increases as the user approaches the edges of their play area.

When the user is at the center of the room, $(x_h, z_h) = (0, 0)$, the offsets are $0$. When the user is at a corner of the room, e.g., at $(x_h, z_h) = (\tfrac{width}{2}, \tfrac{length}{2})$, the offsets become half the noise added to each room dimension $(\tfrac{\mathrm{Noise}_x}{2}, \tfrac{\mathrm{Noise}_z}{2})$. Consequently, the adversary can only collect the noisy room dimensions, e.g., for width: ${x'_h = x_h + \mathrm{offset_x} = \tfrac{width/2}{width}*width' = \tfrac{width'}{2}}$.
Thus, the adversary would only learn a differentially private room dimension from observing $x'_h$ in the range $[0,\tfrac{width'}{2}]$, with the same being true of $\mathit{length}$. Note that offsets added to $x_h$ and $z_h$ are intentionally chosen independently so that the adversary cannot even learn the proportions of the room.


\aptLtoX[graphic=no,type=html]{ \includegraphics{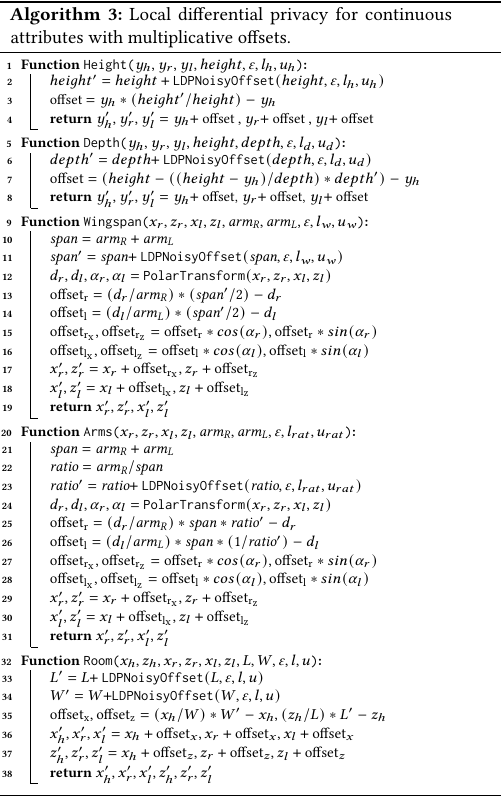} }{ \SetKwComment{Comment}{// }{}
\RestyleAlgo{ruled}
\begin{algorithm}[!ht]
\footnotesize
\DontPrintSemicolon
\caption{Local differential privacy for continuous attributes with multiplicative offsets.}
\label{alg:defenses_continuous_dynamic}


\SetKwFunction{Fsub}{Height}
\SetKwProg{Fn}{Function}{:}{}
\Fn{\Fsub{$y_h, y_r, y_l, height, \varepsilon, l_{h}, u_{h}$}}{
    ${height'=height+\text{\tt LDPNoisyOffset}(height, \varepsilon, l_{h}, u_{h})}$\;
    offset $=y_h*(height'/height) - y_h$\;
    \KwRet  $y'_h, y'_r, y'_l = y_h + $ offset $, y_r + $ offset $, y_l + $ offset 
}

\SetKwFunction{Fsub}{Depth}
\SetKwProg{Fn}{Function}{:}{}
\Fn{\Fsub{$y_h, y_r, y_l, height, depth, \varepsilon, l_{d}, u_{d}$}}{
    $depth' = depth +$ {\tt LDPNoisyOffset}$(depth, \varepsilon, l_{d}, u_{d})$\;
    ${\mathrm{offset} = \left( height - ((height - y_h)/depth)*depth' \right) - y_h}$\;
    \KwRet $y'_h, y'_r, y'_l = y_h + $ offset, $y_r + $ offset, $y_l + $ offset\;
}
\SetKwFunction{Fsub}{Wingspan}
\SetKwProg{Fn}{Function}{:}{}
\Fn{\Fsub{$x_r, z_r, x_l, z_l, \mathit{arm_R}, \mathit{arm_L}, \varepsilon, l_{w}, u_{w}$}}{
    $\mathit{span} = \mathit{arm_R} + \mathit{arm_L}$\;
    $\mathit{span}' = \mathit{span} +$ {\tt LDPNoisyOffset}$(\mathit{span}, \varepsilon, l_{w}, u_{w})$\;
    
    $d_r, d_l, \alpha_r, \alpha_l = $ {\tt PolarTransform}$(x_r, z_r, x_l, z_l)$\;
    
    $\mathrm{offset_r} = (d_r/\mathit{arm_R})* (\mathit{span}'/2) - d_r$\;
    $\mathrm{offset_l} =  (d_l/\mathit{arm_L})*(\mathit{span}'/2) - d_l$
    
    $\mathrm{offset_{r_x}}, \mathrm{offset_{r_z}} = \mathrm{offset_r}*cos(\alpha_r), \mathrm{offset_r}*sin(\alpha_r)$\;
    
    $\mathrm{offset_{l_x}}, \mathrm{offset_{l_z}} = \mathrm{offset_l}*cos(\alpha_l), \mathrm{offset_l}*sin(\alpha_l)$

    $x'_r, z'_r = x_r+\mathrm{offset_{r_x}}, z_r+\mathrm{offset_{r_z}}$\;
    
    $x'_l, z'_l = x_l+\mathrm{offset_{l_x}}, z_l+\mathrm{offset_{l_z}}$
    
    \KwRet  $x'_r, z'_r, x'_l, z'_l$\;
}

\SetKwFunction{Fsub}{Arms}
\SetKwProg{Fn}{Function}{:}{}
\Fn{\Fsub{$x_r, z_r, x_l, z_l, \mathit{arm_R}, \mathit{arm_L}, \varepsilon, l_{rat}, u_{rat}$}}{
    $\mathit{span} = \mathit{arm_R} + \mathit{arm_L}$\;
    $\mathit{ratio} = \mathit{arm_R}/\mathit{span}$\;
    $\mathit{ratio}' = \mathit{ratio} +$ {\tt LDPNoisyOffset}$(\mathit{ratio}, \varepsilon, l_{rat}, u_{rat})$\;
    
    $d_r, d_l, \alpha_r, \alpha_l = $ {\tt PolarTransform}$(x_r, z_r, x_l, z_l)$\;
    
    $\mathrm{offset_r} =  (d_r/\mathit{arm_R})* \mathit{span}* \mathit{ratio}' - d_r$\;
    
    $\mathrm{offset_l} = (d_l/\mathit{arm_L})* \mathit{span}* (1/\mathit{ratio}') - d_l$
    
    $\mathrm{offset_{r_x}}, \mathrm{offset_{r_z}} = \mathrm{offset_r}*cos(\alpha_r), \mathrm{offset_r}*sin(\alpha_r)$\;

    $\mathrm{offset_{l_x}}, \mathrm{offset_{l_z}} = \mathrm{offset_l}*cos(\alpha_l), \mathrm{offset_l}*sin(\alpha_l)$

    $x'_r, z'_r = x_r+\mathrm{offset_{r_x}}, z_r+\mathrm{offset_{r_z}}$\;
    
    $x'_l, z'_l = x_l+\mathrm{offset_{l_x}}, z_l+\mathrm{offset_{l_z}}$
    
    \KwRet  $x'_r, z'_r, x'_l, z'_l$\;
}

\SetKwFunction{Fsub}{Room}
\SetKwProg{Fn}{Function}{:}{}
\Fn{\Fsub{$x_h, z_h, x_r, z_r, x_l, z_l, L, W, \varepsilon, l, u$}}{
    $L' = L + $ {\tt LDPNoisyOffset}$(L, \varepsilon, l, u)$\;
    $W' = W + ${\tt LDPNoisyOffset}$(W, \varepsilon, l, u)$\;
    $\mathrm{offset_x}, \mathrm{offset_z}= (x_h/W)*W' - x_h, (z_h/L)*L' - z_h$\;
    $x'_h, x'_r, x'_l = x_h + \mathrm{offset}_x, x_r + \mathrm{offset}_x, x_l + \mathrm{offset}_x$\;
    $z'_h, z'_r, z'_l = x_h + \mathrm{offset}_z, z_r + \mathrm{offset}_z, z_l + \mathrm{offset}_z$\;
    \KwRet  $x'_h, x'_r, x'_l, z'_h, z'_r, z'_l$\;
}
\end{algorithm}}

\bigskip

\noindent{\textit{Security Arguments}.} We conclude by arguing why the multiplicative offset approach maintains differential privacy, emphasizing that applying a fixed $\mathit{offset}$ multiplicatively is very different from re-sampling the random $\mathit{offset}$ value.

\begin{proposition}
\label{proposition:DP_dynamic_offset}
\textit{Given an single individual's ground truth value $v\in[l, u]$ collected locally once, where $l$ and $u$ are the lower and upper bounds of possible values of $v$, and an offset $\mathrm{N}$ sampled once from a differentially private distribution, broadcasting any $v' = \tfrac{w}{v} (v + \mathrm{N})$ to a server protects $v$ with differential privacy, where $w\in[0, v]$ is a real-time value continuously generated locally.}
\end{proposition}
\textit{Proof:} 
Firstly, an adversary cannot learn the sensitive value from the ratio $\tfrac{w}{v} \in [0,1]$ without knowing $w$. Thus, an adversary can only learn $v + \mathrm{N}$ from the possible stream of broadcasted values $v'=\{0, ..., v + \mathrm{N}\}$ sent to the server. Given that $\mathrm{N}$ is sampled from a differentially private distribution s.t. $v+N$ is centered around $v$, $v + \mathrm{N}$ is immune to post-processing and is thus differentially private~\cite{dworkalgorithmic2013}.
\hfill \qedsymbol{}

\bigskip

\noindent To provide a concrete example, consider again the attribute of \emph{height}: $v=height, v'=height+\mathit{offset}, w=y_h$. Given that $height'$ is differentially private, an adversary who does not know the user's current $y_h$ value (between $0$ and $height$) will only be able to observe the current $y'_h$ value (between $0$ and $height'$), which cannot be used to find $height$.


\subsection{Binary Attributes}
\label{subsec:binary_attributes}

We now switch our focus to attributes like \emph{handedness} which can be represented as Boolean variables. For such attributes, we deploy the {\tt RandomizedResponse} function of Alg.~\ref{alg:defenses}. If randomized response suggests an untruthful response, the user's virtual avatar is mirrored for other users, as is their view of the virtual world.
While the user can still interact with the world and other avatars normally, we found that this approach comes at the cost of all text appearing to be backwards absent any special corrective measures.

\smallskip

\noindent{\textit{Handedness}.}
An adversary may observe a user's behavior, e.g., which hand they use to interact with virtual objects, to determine their handedness over time.
Mirroring the user's avatar randomly on each VR session obfuscates handedness.

\smallskip

\noindent{\textit{Arm Length Asymmetry}.}
Using a mirrored avatar also provides plausible protection against adversaries observing which arm is longer; however, there is a large degree of overlap between this defense and that of \emph{arm length ratio}.
\vspace{-0.3em}

\subsection{Network Communication Attributes}
\label{subsec:network_attributes}

Finally, we turn our attention to network-layer attributes, namely \emph{latency}, which can reveal \emph{geolocation} via multilateration, and \emph{throughput}, which can reveal the \emph{VR device model}.
Such attributes are extremely difficult to protect with differential privacy due to their one-way boundedness; for example, while we can add artificial delay to increase perceived latency, there is no way to decrease the latency of a system below its intrinsic value, which would be necessary to provide differential privacy based on the ground truth. Instead, we resort to clamping, which has the effect of grouping observed attribute values into distinct clusters that effectively anonymize users within their cluster.
The defenses of this section are not intended to be a primary contribution of our paper, but are a necessary component of a complete VR privacy solution.

\smallskip

\noindent{\textit{Geolocation}.} A server attacker can observe the round-trip delay of a signal traveling between a VR client device and multiple servers to determine a user's location via multilateration (hyperbolic positioning). Furthermore, prior works suggest a user attacker can also use the round-trip delay of the target's audio signal as a proxy for latency.
In response, our defense clamps the latency of all broadcasted signals to a fixed round-trip delay by artificially delaying each packet. Due to the sensitivity of hyperbolic positioning, even a $1$~ms offset can skew the adversaries' prediction by $\approx 300$~km.

\smallskip

\noindent{\textit{Reaction Time}.}
Likewise, adversaries can measure a user's reaction time by timing the delay between a stimulus (e.g., a visual or audio cue) and the user's response.
In addition to being a further identifying metric, reaction time is also highly correlated with age~\cite{woodsagerelated2015}.
While the technical defense for reaction time is largely equivalent to that of geolocation, the specific clamping values are different because the sensitivity of the underlying attributes vary greatly. If the defenses for geolocation and reaction are simultaneously enabled, the higher latency clamp should be applied to protect both.

\smallskip


\noindent{\textit{Refresh/Tracking Rate}.} Finally, a server attacker can use the telemetry throughput to ascertain the VR headset's refresh and tracking rate and thus potentially identify the make and model of a user's VR device.
Moreover, user attackers can leverage a VR environment with moving objects that users perceive differently depending on their refresh rates to determine the refresh rate of the VR display.
Thus, our defenses clamp the rate at which the VR device broadcasts its tracked coordinates to obfuscate the true device specifications.

\smallskip

\noindent \textbf{Summary.} While our aim in this section was to be as thorough as possible with regard to covering known VR privacy attacks, we by no means claim to have comprehensively addressed every possible VR privacy threat vector. Instead, we hope to have accomplished two simple goals.
Firstly, we believe the combined defenses of this section are sufficient to significantly hinder attempts to deanonymize users in the metaverse. Within a large enough group of users, adversaries may have to combine dozens of unique attributes to reliably identify individuals; the absence of the low-hanging attributes discussed herein should obstruct their ability to do so.
Secondly, we hope that the attributes covered in this section were diverse enough, and the corresponding defenses flexible enough, to be extended to future VR privacy threats.


\begin{figure*}[!t]
    \centering
    \includegraphics[width=\linewidth]{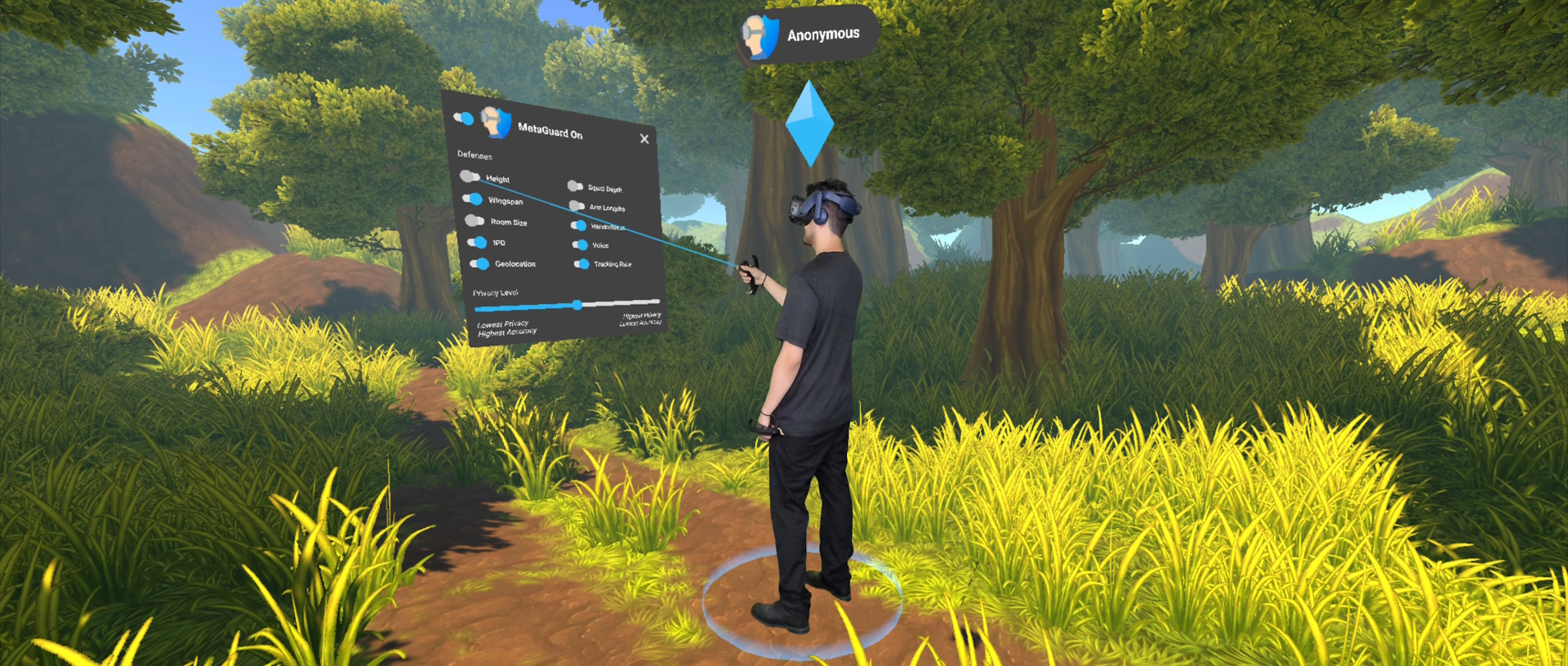}
    \caption{Mixed reality photo of a player using ``MetaGuard,'' our implementation of incognito mode for VR.}
    \label{fig:XR}
\end{figure*}
\section{VR Incognito Mode}
\label{sec:VR_incognito}

In this section, we introduce ``MetaGuard,''\footnote{Short for ``Metaverse Guard.''} our practical implementation of the defenses presented in~\S\ref{sec:Privacy_defenses} and the first known ``incognito mode'' for the metaverse. We built MetaGuard as an open-source Unity (C\#) plugin that can easily be patched into virtually any VR application using MelonLoader \cite{melonloader}.\footnote{Unlike mobile apps, desktop VR apps can be modified by end users.} We begin by describing the options and interface made available to MetaGuard users. We then discuss our choice of DP parameters ($\varepsilon$, bounds, etc.) and outline how MetaGuard calibrates noise to each user. Finally, we describe the concrete game object transformations applied to the virtual world to implement the defenses of \S\ref{sec:Privacy_defenses}. Fig.~\ref{fig:XR} shows a mixed reality photo of a player using the MetaGuard VR plugin within a VR game.

\subsection{Settings \& User Interface}
\label{subsec:user_interface}

\begin{figure}[H]
    \centering
    \includegraphics[width=\linewidth]{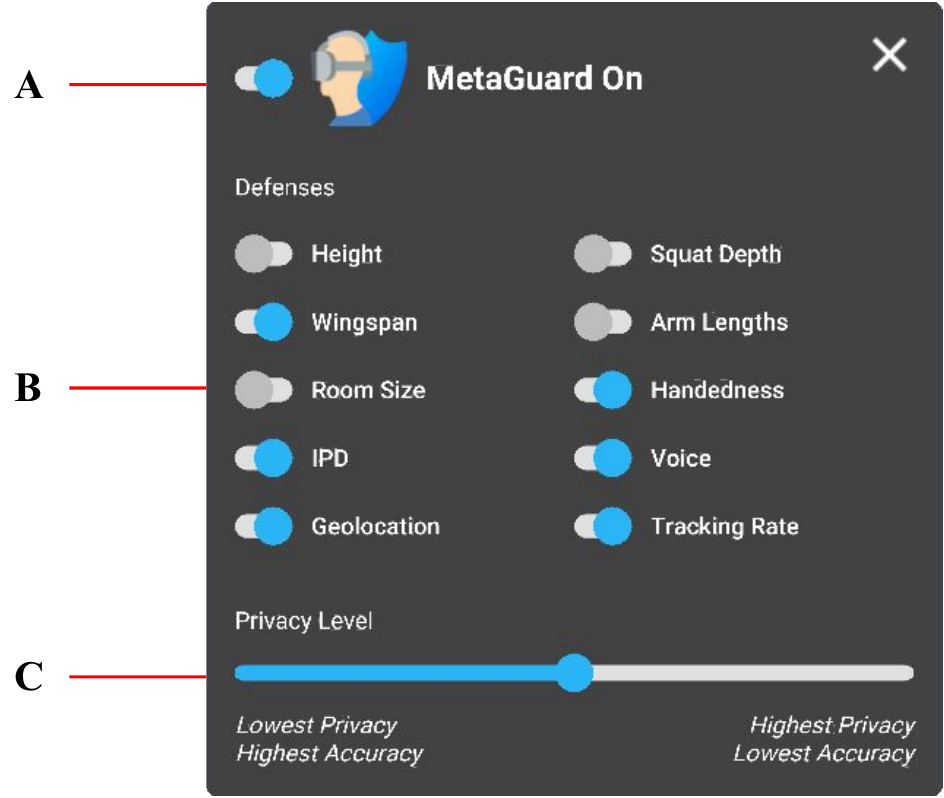}
    \caption{VR user interface of MetaGuard plugin.}
    \label{fig:UI}
\end{figure}

The main objective of MetaGuard is to protect VR user privacy while minimizing usability impact. The flexible interface of MetaGuard (shown in Fig. \ref{fig:UI}) reflects this goal, allowing users to tune the defense profile according to their preferences and to the needs of the particular VR application in use. Specifically, we expose the following options:


\medskip

\noindent \textbf{(A) Master Toggle.} The prominent master switch allows users to ``go incognito'' at the press of a button, with safe defaults that invite (but don't require) further customization.

\medskip

\noindent \textbf{(B) Feature Toggles.} The feature switches allow users to toggle individual defenses according to their needs; e.g., in a game like Beat Saber~\cite{beatSaber}, users may wish to disable defenses that interfere with gameplay (i.e., wingspan and arm lengths), while keeping the other defenses enabled.

\noindent \textbf{(C) Privacy Slider.} Lastly, we present users with a ``privacy level'' slider that adjusts the privacy parameter ($\varepsilon$) for each defense, allowing users to dynamically adjust the inherent trade-off between privacy and accuracy when using the defenses of \S\ref{sec:Privacy_defenses}. Users can choose from the following options, which we generally refer to simply as the ``low,'' ``medium,'' and ``high'' privacy settings:


\begin{itemize}[leftmargin=*]
    \itemsep 0em
    \item \textbf{High Privacy}, intended for virtual telepresence applications such as VRChat~\cite{vrchat} and others~\cite{altvr, metahorizon}. 
    \item \textbf{Balanced}, intended for casual gaming applications, such as virtual board games requiring some dexterity~\cite{tabletop}.
    \item \textbf{High Accuracy}, intended for noise-sensitive competitive gaming applications~\cite{raceroom} such as Beat Saber~\cite{beatSaber}.
\end{itemize}



\subsection{Selecting Epsilon Values \& Attribute Bounds}
\label{subsec:epsilons}

As discussed in \S\ref{subsec:differential_privacy}, the level of privacy provided by the defenses of \S\ref{sec:Privacy_defenses} depends on the appropriate selection of DP parameters, namely $\varepsilon$, $\Delta$, and attribute bounds. Although our approach in MetaGuard is to allow users to adjust the privacy parameter ($\varepsilon$) according to their preferences, we must nevertheless translate the semantic settings of ``low,`` ``medium,`` and ``high`` privacy into concrete $\varepsilon$-values, noting that a given privacy level may translate to a different $\varepsilon$-value for each attribute depending on its sensitivity to noise. Furthermore, the specific lower bound ($l$) and upper bound ($u$) of each attribute (and thus $\Delta=|u-l|$) must be determined in order to use the Bounded Laplace mechanism. This section outlines our method of selecting these values, with the results shown in Tab. \ref{tab:epsilons}.

\subsection*{Selecting $\varepsilon$-Values \& Clamps}

\noindent \textbf{Continuous Anthropometrics.}
We conducted a small empirical analysis using the primary authors of this paper\footnote{The authors include one novice VR user and one expert.} to select appropriate $\varepsilon$-values for each of the continuous anthropometric attributes at each privacy level.
We began by selecting three VR applications (VRChat~\cite{vrchat}, Tabletop Simulator~\cite{tabletop}, and Beat Saber~\cite{beatSaber}) that represent the most popular examples of the intended use cases for the high, medium, and low privacy modes respectively.
We then tested a wide range of $\varepsilon$-values for each attribute in each application while monitoring their effect on usability. For example, in Beat Saber, we had both a novice and expert-level player complete the same challenges at different $\varepsilon$-values to evaluate the impact of noise on in-game performance. By contrast, in VRChat, we were simply interested in the impact of noise on the ability to hold a conversation (e.g., to maintain virtual ``eye contact'').

\vspace{-0.75em}

\begin{figure}[h]
    \centering
    \includegraphics[width=\linewidth]{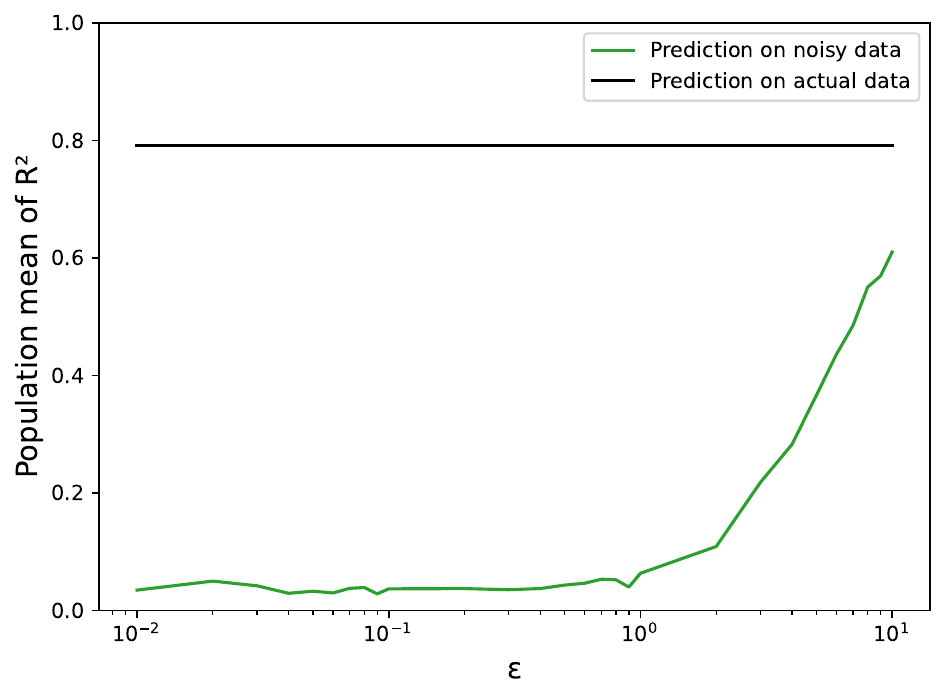}
    \vspace*{-2.5em}
    \caption{Coefficients of determination of height from predictions on actual vs. noisy data as $\varepsilon$ increases.}
    \label{fig:epsselect}
\end{figure}

\vspace{-0.75em}

Next, we analyzed the concrete privacy impact of candidate $\varepsilon$ choices by simulating attackers at a variety of $\varepsilon$-values. For example, Fig. \ref{fig:epsselect} illustrates that for the height attribute, the vast majority of privacy benefit is already realized at $\varepsilon=1$. We combined these results with the findings of our usability analysis to produce the final $\varepsilon$-values shown in Tab. \ref{tab:epsilons} according to the appropriate balance of privacy and usability for the intended use of each level.

\noindent \textbf{Binary Anthropometrics.}
For attributes where the defenses of \S\ref{sec:Privacy_defenses} suggest the use of randomized response, we selected $\varepsilon$-values such that the corresponding prediction accuracy was degraded by $15$\%, $50$\% and $85$\% at the low, medium, and high privacy levels.

\medskip

\noindent \textbf{Voice.} Although technically a continuous anthropometric, vocal frequency cannot be calibrated via playthroughs due to the lack of a tangible impact on gameplay performance.
Instead, we selected $\varepsilon$-values which degraded inference of gender by roughly $25$\%, $50$\% and $75$\% at the low, medium, and high privacy levels respectively. 

\medskip

\noindent \textbf{Clamps.} Finally, for attributes where the corresponding defense of \S\ref{sec:Privacy_defenses} suggests clamping, we chose clamp values which have the effect of anonymizing users within progressively larger groups. For example, for refresh/tracking rate, we selected clamps which hide users within the set of high ($90$Hz~\cite{highHz}), medium ($72$Hz~\cite{medHz}), and low ($60$Hz~\cite{lowHz}) fidelity VR devices. For the latency-related attributes, we selected values below the perceptible $100$ms threshold~\cite{100ms1, 100ms2, 100ms3} that significantly decreased prediction accuracy.

\medskip

\subsection*{Selecting Attribute Bounds}
Finally, beyond $\varepsilon$, the Bounded Laplace mechanism also requires attribute bounds to constrain the outputs to semantically consistent values. We used public datasets to obtain the $95$th percentile bounds for anthropometric measurements \cite{heightpercentiles, dodgson2004variation, repreferences2012, sarmacorrelation2020}; our use of local DP causes $\Delta$ to reflect the full range of possible values. For room size, we extracted the bounds from official VR setup specifications~\cite{roomsizes}. We list the bounds and corresponding references in Tab.~\ref{tab:epsilons}.

\begin{table}[h]
\resizebox{\columnwidth}{!}{%
\begin{tabular}{|l|ll|lll|}
\hline
\multirow{2}{*}{\textbf{Data Point}} & \multicolumn{2}{c|}{\textbf{Bounds}} & \multicolumn{3}{c|}{\textbf{Privacy Levels}} \\ \cline{2-6} 
 & \multicolumn{1}{l|}{\textbf{Lower}} & \textbf{Upper} & \multicolumn{1}{l|}{\textbf{Low}} & \multicolumn{1}{l|}{\textbf{Medium}} & \textbf{High} \\ \hline
 
Height~\cite{heightpercentiles} & \multicolumn{1}{l|}{1.496m} & 1.826m & \multicolumn{1}{l|}{$\epsilon$=5} & \multicolumn{1}{l|}{$\epsilon$=3} & $\epsilon$=1 \\ \hline

IPD~\cite{dodgson2004variation} & \multicolumn{1}{l|}{55.696mm} & 71.024mm & \multicolumn{1}{l|}{$\epsilon$=5} & \multicolumn{1}{l|}{$\epsilon$=3} & $\epsilon$=1 \\ \hline

Voice Pitch~\cite{repreferences2012} & \multicolumn{1}{l|}{85 Hz} & 255 Hz & \multicolumn{1}{l|}{$\epsilon$=6} & \multicolumn{1}{l|}{$\epsilon$=1} & $\epsilon$=0.1 \\ \hline

Squat Depth~\cite{nair2022exploring} & \multicolumn{1}{l|}{0m} & 0.913m & \multicolumn{1}{l|}{$\epsilon$=5} & \multicolumn{1}{l|}{$\epsilon$=3} & $\epsilon$=1 \\ \hline

Wingspan~\cite{sarmacorrelation2020} & \multicolumn{1}{l|}{1.556m} & 1.899m & \multicolumn{1}{l|}{$\epsilon$=3} & \multicolumn{1}{l|}{$\epsilon$=1} & $\epsilon$=0.5 \\ \hline

Arm Ratio~\cite{nair2022exploring} & \multicolumn{1}{l|}{0.95} & 1.05 & \multicolumn{1}{l|}{$\epsilon$=3} & \multicolumn{1}{l|}{$\epsilon$=1} & $\epsilon$=0.5 \\ \hline

Room Size~\cite{roomsizes} & \multicolumn{1}{l|}{0m} & 5m & \multicolumn{1}{l|}{$\epsilon$=3} & \multicolumn{1}{l|}{$\epsilon$=1} & $\epsilon$=0.1 \\ \hline

Handedness & \multicolumn{1}{l|}{0} & 1 & \multicolumn{1}{l|}{$\epsilon$=1.28} & \multicolumn{1}{l|}{$\epsilon$=0.88} & $\epsilon$=0.73 \\ \hline

Latency (Geolocation) & \multicolumn{2}{l|}{Clamped} & \multicolumn{1}{l|}{25ms} & \multicolumn{1}{l|}{30ms} & 50ms \\ \hline

Reaction Time & \multicolumn{2}{l|}{Clamped} & \multicolumn{1}{l|}{10ms} & \multicolumn{1}{l|}{20ms} & 100ms \\ \hline

Refresh/Tracking Rate & \multicolumn{2}{l|}{Clamped} & \multicolumn{1}{l|}{90 Hz} & \multicolumn{1}{l|}{72 Hz} & 60 Hz \\ \hline

\end{tabular}%
}
\caption{Selected $\varepsilon$, clamps, and attribute bound values.}
\label{tab:epsilons}
\end{table}

\vspace{-2em}

We emphasize that the sole purpose of our informal experimentation in this section is to set a reasonable range of $\varepsilon$-values that cover a variety of VR use cases. 
Given the lack of consensus on a formal method for selecting DP parameters~\cite{dworkdifferential2019}, our choices simply serve to establish a plausible spectrum of $\varepsilon$-values corresponding to our perceived boundaries of the privacy-usability trade-off.
The power to select exactly which point on this spectrum is best suited for a particular application remains with the end user.

\subsection{Rerandomization \& Linkability}
By default, we suggest randomly resampling offset values according to the algorithms of \S\ref{sec:Privacy_defenses} at the start of each session. Assuming that MetaGuard users cannot be linked across sessions, adversaries will be unable to aggregate measurements across multiple sessions to obtain user data. Alternatively, one-time randomization can be used, allowing linkability but assuring no attribute leakage occurs.


\subsection{Calibration \& Noise Centering}
\label{subsec:noise_calibration}

One final parameter is required to successfully implement the continuous attribute defenses of \S\ref{sec:Privacy_defenses}: the ground truth attribute values of the end user. Centering the Laplacian noise distribution around the ground truth attribute values of the current user has the effect of minimizing noise for as many users as possible, particularly those who are outliers, thus achieving theoretically optimal usability.

\begin{figure}[h]
    \centering
    \includegraphics[width=\linewidth]{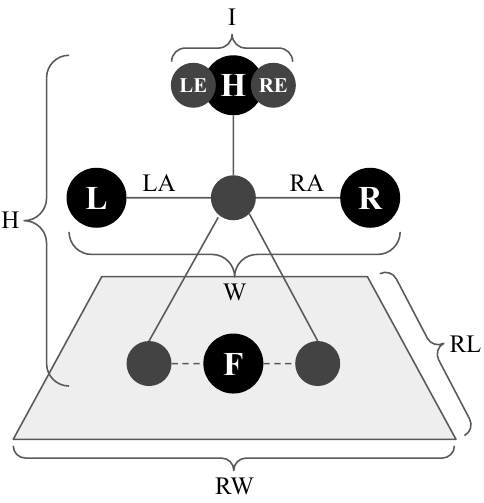}
    \caption{Instantaneous calibration of ground truth for height (H), left arm (LA), right arm (RA), wingspan (W), IPD (I), room width (RW), and room length (RL), using head (H), floor (F), left/right controllers (L/R); figure not to scale.}
    \label{fig:gt}
\end{figure}

To achieve this, the MetaGuard extension calculates instantaneous ground truth estimates upon instantiation using the method shown in Fig. \ref{fig:gt}.
Specifically, the OpenVR API~\cite{unitydocs} provides MetaGuard with one-time snapshot locations of the user's head, left and right eyes, left and right hands, and a plane representing the play area.
Estimates for the ground truth values of height, wingspan, IPD, room size, and left and right arm lengths can then be derived from these measurements.
We note that the privacy of MetaGuard is not dependent on the accuracy of the ground truth estimates, which exist only to ensure that the added noise is not more than the level necessary to protect a given user.

\subsection{Defense Implementation}
\label{subsec:defense_implementation}


We now finally provide a complete description of our ``VR Incognito Mode'' system for implementing the defenses of \S\ref{sec:Privacy_defenses} in light of the interface, $\varepsilon$-values, bounds, and calibration procedures described above. Our implementation follows two phases: a \emph{setup phase}, which executes exactly once on the frame when a defense is enabled, and an \emph{update phase}, which executes every frame thereafter.\medskip

\begin{figure}[h]
    \centering
    \includegraphics[width=\linewidth]{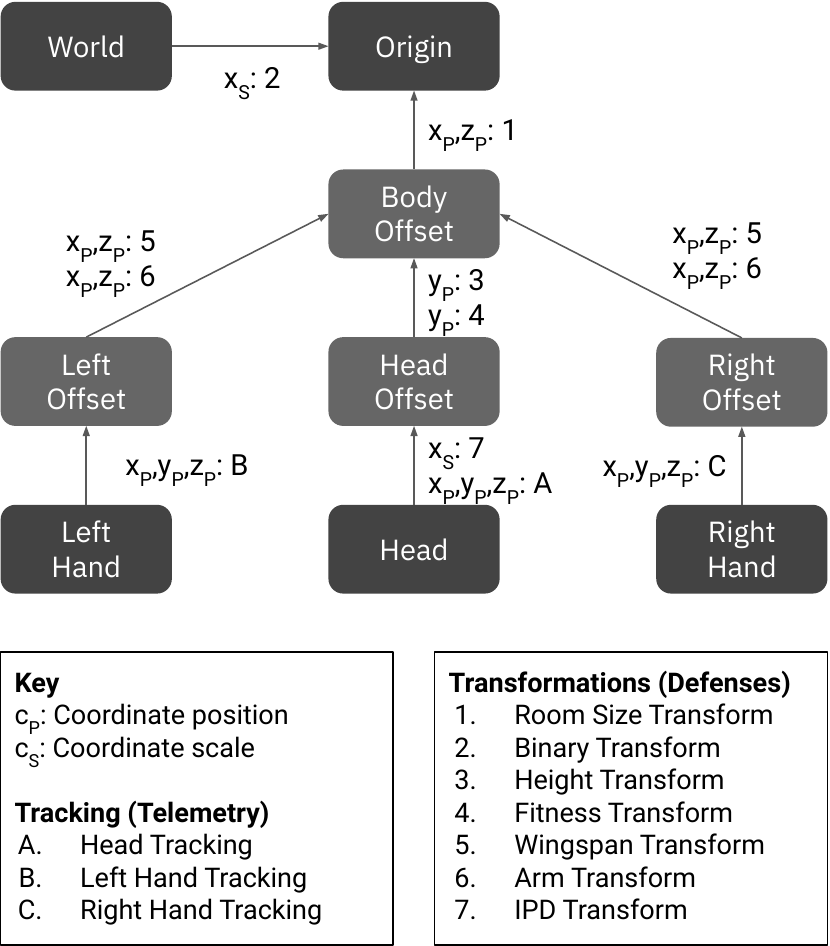}
    \caption{Game object hierarchy with existing (dark grey) and inserted (light grey) game objects, and coordinate transformations used to implement VR Incognito Mode defenses.}
    \label{fig:implementation}
\end{figure}

\vspace{-0.6em}

\noindent{\textbf{Setup Phase}.} When a defense is first enabled, MetaGuard uses the calibration procedures of~\S\ref{subsec:noise_calibration} to estimate the ground truth attribute values of the user.
These values are then used in combination with the $\varepsilon$-values and bounds of~\S\ref{subsec:epsilons} to calculate noisy offsets corresponding to each privacy level using the methods outlined in \S\ref{sec:Privacy_defenses}, and are then immediately discarded from program memory (with only offsets retained) so as to minimize the chance of unintentional data leakage.
By default, the Unity game engine uses telemetry data from OpenVR~\cite{openvr} to position game objects within a virtual environment, which are then manipulated by a VR application.
During the setup phase, the system modifies the game object hierarchy by inserting intermediate ``offset" objects as shown in Fig.~\ref{fig:implementation}.

\medskip

\noindent{\textbf{Update Phase}.} During the update phase, the system first checks which defenses the user has enabled in the interface (see \S\ref{subsec:user_interface}).
For all disabled attributes, the corresponding offset transformations in the game object hierarchy (as shown in Fig.~\ref{fig:implementation}) are set to the identity matrix.
For each enabled feature, the system implements the corresponding defense of~\S\ref{sec:Privacy_defenses} by fetching the noisy attribute value calculated during the setup phase for the currently-selected privacy level and enabling the relevant coordinate transformation on the inserted offset objects such that the observable attribute value matches the noisy attribute value.
Specifically, Fig.~\ref{fig:implementation} illustrates how the position of each game object is defined with respect to another object in the hierarchy, and how the defenses modify the relative position or scale of each object with respect to its parent.

\section{System Evaluation \& Results}
\label{sec:results}

In this section, we demonstrate the effectiveness of the defenses introduced in~\S\ref{sec:Privacy_defenses} by evaluating their impact on the accuracy of a theoretical attacker. To do so, we faithfully replicated the attacks of the TTI~\cite{millerpersonal2020}, MetaData~\cite{nair2022exploring}, and 50k~\cite{nair2023unique} studies to measure their accuracy both with no defenses and with the MetaGuard extension at the low, medium, and high privacy levels. The results of this evaluation are summarized in Tab. \ref{tab:results} of \S\ref{app:results}. The presented accuracy values represent what a server attacker could achieve, and also provide an upper bound for the capabilities of user attackers.

\subsection{Evaluation Method}
\label{subsec:method}

We obtained from the original authors anonymized frame-by-frame telemetry data recordings of the $511$ users from the TTI~\cite{millerpersonal2020} study, $30$ users from the MetaData~\cite{nair2022exploring} study, and $55,541$ users from the 50k~\cite{nair2023unique} study.
Using this data, we could virtually ``replay'' the original sessions exactly as they occurred, and were able to reproduce the identification and inference attacks described in the original studies with nearly identical results. Next, we repeated this process for each session with MetaGuard enabled at the low, medium, and high privacy levels. The resulting decrease in attack accuracy for each attribute at each privacy level is shown in \S\ref{app:results}.

To emulate a realistic metaverse threat environment, we streamed telemetry data from the client to a remote game server via a WebSocket. The MetaGuard extension was allowed to clamp the bandwidth and latency of this data stream as discussed in \S\ref{sec:Privacy_defenses}. The network-related attacks were then run on the server side.

Beyond the attacks which deterministically harvest sensitive data attributes, all three studies use machine learning to identify users or profile their demographics. We used sklearn to replicate the published methods as closely as possible, using the same model types and parameters as in the original papers. Once again, we replicated the original results with similar accuracy, with the decrease in identification corresponding to the use of the low, medium, and high privacy levels of MetaGuard being shown in Tab. \ref{tab:results}C of \S\ref{app:results}.


\subsection{Ethical Considerations}
\label{subsec:ethical_considerations}

Other than the $\varepsilon$-calibration effort described in \S\ref{subsec:epsilons}, which was performed by the authors, this paper does not involve any original research with human subjects. Instead, our results rely on the replication of prior studies using anonymous data obtained either from public online repositories or directly from the authors of those studies.
We verified that all original studies from which we obtained data were non-deceptive and were each subject to individual ethics review processes by OHRP-registered institutional review boards.
Furthermore, the informed consent documents of those studies explicitly included permission to re-use \mbox{collected} data for follow-up studies, and we strictly followed the data handling requirements of the original consent documentation, such as the promise to only publish statistical aggregates rather than individual data points.

\subsection{Primary \& Secondary Attributes}
\label{subsec:primary_secondary_attributes}

\noindent \textbf{Continuous Anthropometrics.}
Tab.~\ref{tab:results}A shows that our defenses effectively reduce the coefficients of determination to values below $0.5$ for the targeted continuous attributes.
We found that physical fitness (squat depth) is the most challenging attribute to protect while preserving user experience, as it shows the smallest drops in prediction accuracy. The remaining attributes show significant decreases in attack \mbox{accuracy} even at the low privacy level: IPD ($-67.53\%$), room size ($-55.89\%$ within $2$m²), wingspan ($-33.07\%$ within $7$~cm) and height ($-16.93\%$ within $5$~cm). \smallskip

\noindent \textbf{Binary Anthropometrics.}
An advantage of the randomized response technique is precise control over attacker accuracy levels by choosing the values of $\varepsilon$.
Unsurprisingly, the prediction accuracy of handedness (92.5\%, 75\%, and 57.5\% for the low, medium, and high privacy levels) corresponded to the chosen $\varepsilon$-values. \smallskip

\noindent \textbf{Network Attributes.}
The prediction accuracy of the attributes dependent on latency and throughput dramatically dropped thanks to clamping (except for reaction time, which showed a modest accuracy drop of $8.3\%$ at a low privacy level). 
Altogether, the low accuracy of these predictions significantly impedes the ability of adversaries to determine which VR device an individual is using. 

\subsection{Inferred Attributes}
\label{subsec:inferred_attributes}

The machine learning models of the MetaData study primarily use the attributes discussed above as model inputs to infer demographics. Clearly, the reduction in accuracy of these primary attributes will have a negative impact on the accuracy of inferences based on them; nonetheless, we ran the models on the noisy attributes to quantify this impact.
The results show significant accuracy drops in predicting gender ($-23.5\%$), age ($-58.25\%$), ethnicity ($-48.75\%$), and income ($-73.85\%$), even at the lowest privacy setting.
Most importantly, the three identification models simulating an attacker identifying a user amongst a group all had a significant drop in accuracy (see Tab.~\ref{tab:results}C); thus, MetaGuard empirically succeeds at its primary goal of preventing users from being deanonymized.
\section{Discussion}
\label{sec:Discussion}

In this study, we set out to design, implement, and evaluate a comprehensive suite of VR privacy defenses to protect VR users against a wide range of known attacks. In the absence of any defenses, these attacks demonstrated the ability to not only infer specific sensitive attributes, but also to combine these attributes to infer demographics and even deanonymize users entirely.

Through our evaluation of MetaGuard, our practical implementation of a ``VR incognito mode'' plugin, we have demonstrated that $\varepsilon$-differential privacy can pose an effective countermeasure to such attacks.
Our results show a considerable accuracy reduction in the identification and profiling of users using real VR user data from 56,082 participants across three popular VR privacy studies.
By evaluating our system using telemetry data from these existing studies, we were able to independently measure the performance of each defense at each supported privacy level, a feat that would otherwise have required an infeasible number of laboratory trials.

MetaGuard allows users to ``go incognito'' by randomizing their fictitious measurements, such as height and wingspan, at the start of each new session, thus thwarting cross-session likability. Alternatively, if users do not mind being linked across sessions, they do not need to re-randomize their fictitious measurements between sessions, allowing adversaries to track them across sessions without revealing their true attribute values in the process.

Our use of bounded Laplacian noise allows us to achieve a theoretically optimal balance between privacy and usability, minimizing the mean squared tracking error a user is expected to experience for a given privacy level ($\varepsilon$) \cite{boundedtruncatedIBM, dworkalgorithmic2013}. This, in turn, allows us to leverage homuncular flexibility to implement the defenses in a way that users can rapidly learn to ignore \cite{won2015homuncular, abtahi2022beyond}. For example, the average wingspan offset at the medium privacy level is 4.5~cm, which is well within the range that VR users can flexibly adapt to \cite{poupyrev1996go}. Even those transformations which do not directly affect the player model can be thought of as equivalent to body modifications. For example, room size is not necessarily implemented as a body manipulation, but changing the room-to-avatar ratio can be thought of as equivalent to changing the size of the entire avatar and thereby hiding the relative size of the room. As such, we expect homuncular flexibility to be applicable to such transformations as well.


Overall, MetaGuard constitutes the first attempt at producing a privacy-preserving ``incognito mode'' solution for VR. Grounded in theoretical privacy, and demonstrated using thorough empirical evaluation, we aim to provide a solid foundation for future work in this area.
The importance of privacy-enhancing software like MetaGuard will become more pronounced as current market trends make virtual reality increasingly ubiquitous and shape the next generation of the social internet, the so-called ``metaverse''~\cite{metaverseintro, Blackrockmeta, Morganmeta}.
As it stands, VR device manufacturers have been observed selling VR hardware at losses of up to \$$10$~billion per year~\cite{publisheddespite2022}, presumably with the goal of recouping this investment through software-based after-sale revenue, such as via targeted advertisement~\cite{surveillanceads, usenixprivacysurvey}.

But despite using the terms ``attacker,'' and ``adversary'' throughout our writing, it's likely that such actions would in practice be entirely above board, with users agreeing (knowingly or otherwise) to have their data collected. It is more important than ever to give users the ability to protect their data through technological means, independent of any warranted data privacy regulations, in a way that is as easy to use as the privacy tools used on the web today.

\smallskip

\noindent \textbf{Limitations.}
Our decision to base our evaluation on data from prior studies means that we inherit the biases of the original studies. In particular, the test subjects of the studies from which our data is derived were not perfectly representative of the general population of VR users.
While our evaluation method does precisely replicate the telemetry stream that would have been generated by the original participants were they using the MetaGuard extension, it does so under the assumption that their use of MetaGuard would not have changed their behavior. The accuracy of MetaGuard could be somewhat diminished if it turns out that users modify their behavior to compensate for the added noise.
Further, our study considers a limited set of data attributes, which may not be comprehensive with respect to the attributes inferable in VR. MetaGuard may not be effective at protecting attributes beyond those that we directly considered.
Finally, the mean-squared-error definition of ``usability'' by which our system is theoretically optimal may in some cases fail to align with the true user experience in VR.

\smallskip

\noindent \textbf{Future Work.}
Lacking access to VR device firmware, we implemented the MetaGuard extension described in this paper at the client software layer, providing an effective defense against server and user attackers.
In future work, we believe the same defenses could be easily applied at the firmware level, allowing data to also be protected from client attackers.
However, protecting data from hardware or firmware-level adversaries will likely require entirely different methods to the ones presented in this paper.

While our aim in this paper was to be as comprehensive as possible when addressing VR privacy attacks, there were a few niche VR hardware features that we specifically excluded. Future systems could extend the techniques of this paper to less common VR accessories, such as pupil tracking and full-body tracking systems, that we did not address in this work. Moreover, we think it is necessary to enlarge the body of known VR privacy attack vectors, and we hope the framework of the MetaGuard extension is modular enough to support the implementation of their corresponding defenses.

An important aspect of the MetaGuard system is the ability for users to toggle individual VR defenses according to the requirements of the application being used. While this process is entirely manual in our implementation, in the future, the ``incognito mode'' system could be configured to automatically profile VR applications and determine which defenses are appropriate for a given scenario.
Furthermore, the application could incorporate the differential privacy concept of a ``privacy budget,'' adding more noise to enabled attributes to compensate for the privacy loss of disabled attributes and maintain the same level of overall anonymity.
Our method of selecting $\varepsilon$-values was somewhat informal, in part due to the lack of a quantitative metric of usability impact for noise in VR. Therefore, we look forward to future work that performs user studies to rigorously quantify the impact of adding noise to various attributes on the VR user experience, so as to better shed light on the costs vs. benefits of noisy mechanisms like differential privacy in VR.

Finally, there are methods other than differential privacy that, while relinquishing the provability of our approach, may produce a better experience for the end user. In the future, we hope to evaluate techniques that utilize machine learning to develop corruption models that hide user data while maintaining functionality.
\section{Related Work}
\label{sec:Related_Work}

We analyzed a large number of VR/AR/XR security and privacy literature reviews \cite{usenixprivacysurvey, VRreviewanony, metaverseprivacy1, privimplicationsVR, privmetaverse, metaverseintro, VRdisclosureinfo, dickbalancing2021, obrolchainconvergence2016, stephensonsoknodate, deguzmansecurity2020} to asses the current state of the art with respect to metaverse privacy attacks and defenses. The variety of attacks mentioned in these works were a major motivation for producing this paper, as discussed in \S\ref{subsec:vr_privacy_attacks}.

With respect to defenses, there are a limited number of studies proposing the use of differential privacy in VR. Related works have primarily focused on using differential privacy to protect eye-tracking data~\cite{VReyetrackpriv, VReyetrackpriv2, VReyetrackpriv3, snowpixeleyetracking, 263891} without regard to other types of VR telemetry. For example, Steil~et~al.~\cite{VReyetrackpriv} and Ao~et~al.~\cite{VReyetrackpriv2} use differential privacy to protect visual attention heatmaps, while Johnn~et~al.~\cite{snowpixeleyetracking} proposes the use of ``snow'' pixels to obscure the iris signal and prevent spoofing while preserving gaze.

A few of the defenses proposed in this paper have also previously been discussed outside the context of VR. For example, Avery et al.~\cite{averyholding2019} discuss defenses against attacks inferring handedness in the context of mobile devices, and Sun~et.~al~\cite{speechprivacyieee} proposed countermeasures to inferring attributes from speech in mobile applications.


In summary, MetaGuard fills an important gap in the VR privacy landscape, not only by being the first to defend various anthropometric, environmental, demographic, and device attributes, but also in general by presenting a comprehensive usable metaverse privacy solution rather than focusing on any one particular data point.
\section{Conclusion}
\label{sec:Conclusion}

In this paper, we have presented the first comprehensive ``incognito mode for VR.'' Specifically, we designed a suite of defenses that quantifiably obfuscate a variety of sensitive user data attributes with $\varepsilon$-differential privacy.
We then implemented these defenses as a universal Unity VR plugin that we call ``MetaGuard.''
Our implementation, which is compatible with a wide range of popular VR applications, gives users the power to ``go incognito'' in the metaverse with a single click, with the flexibility of adjusting the defenses and privacy level as they see fit.

Upon replicating well-known VR privacy attacks using real user data from prior studies, we demonstrated a significant decrease in attacker capabilities across a wide range of metrics. In particular, the ability of an attacker to deanonymize a VR user was degraded by as much as $96.0$\% while using the MetaGuard extension.

Over the course of decades of research in web privacy, private browsing mode has remained amongst the most ubiquitous privacy tools in popular use today. We were inspired by the success of ``incognito mode'' on the web to produce a metaverse equivalent that is just as user-friendly, while serving the same fundamental purpose of helping users remain untraceable across multiple sessions. We hope our open-source MetaGuard plugin and promising results serve as a foundation for other privacy practitioners to continue exploring usable privacy solutions in this important field.


\section*{Availability}
Our GitHub repository \cite{MetaGuardrepo} contains Unity (C\#) scripts implementing defenses, ``incognito
mode'' plugins for VR applications, and local and remote
evaluation scripts: \\

{\url{https://github.com/metaguard/metaguard}}

\begin{acks}
We thank James O'Brien, Bjoern Hartmann, James Smith, Christopher Harth-Kitzerow, Sriram Sridhar, Xiaoyuan Liu, Lun Wang, Louis Rosenberg, and Syomantak Chaudhuri for their feedback.
This work was supported in part by the National Science Foundation, by the National Physical Science Consortium, and by the Fannie and John Hertz Foundation.
Any opinions, findings, and conclusions or recommendations expressed in this material are those of the authors and do not necessarily reflect the views of the supporting entities.
We sincerely thank the ITT and MetaData study participants for making this work possible.
\end{acks}

\bibliographystyle{ACM-Reference-Format}
\bibliography{999_REFS.bib}

\appendix

\onecolumn
\section{Results}
\label{app:results}

\begin{table}[H]
\resizebox{\textwidth}{!}{%
\begin{tabular}{|l|l|l|l|l|l|}
\multicolumn{6}{c}{\textit{Table \ref{tab:results}A: Primary and Secondary Attributes (MetaData \cite{nair2022exploring} Study)}} \\
\hline
\textbf{Attribute} & \textbf{Metric} & \textbf{No Privacy} & \textbf{Low Privacy} & \textbf{Medium Privacy} & \textbf{High Privacy} \\ \hline 
\hline

\textbf{Height} & \begin{tabular}[c]{@{}l@{}}Within 5cm\\ Within 7cm\\ R²\end{tabular} & \begin{tabular}[c]{@{}l@{}}70\%\\ 100\%\\ 0.79\end{tabular} & \begin{tabular}[c]{@{}l@{}}53.07\% ±2.41\% \\ 68.6\% ±2.18\%\\ 0.37 ±0.040\end{tabular} & \begin{tabular}[c]{@{}l@{}}45.00\% ±2.35\% \\ 58.17\% ±2.09\%\\ 0.22 ±0.035\end{tabular} & \begin{tabular}[c]{@{}l@{}}32.63\% ±2.3\% \\ 44.47\% ±2.43\%\\ 0.06 ±0.020\end{tabular} \\ \hline

\textbf{Physical Fitness} & Categorical & 90\% & 86.11\% ±2.65\% & 79.11\% ±2.60\% & 61.56\% ±4.15\% \\ \hline

\textbf{IPD (Vive Pro 2)} & \begin{tabular}[c]{@{}l@{}}Within 0.5mm\\ R²\end{tabular} & \begin{tabular}[c]{@{}l@{}}96\%\\ 0.991\end{tabular} & \begin{tabular}[c]{@{}l@{}}18.53\% ±1.76\%\\ 0.399 ±0.041\end{tabular} & \begin{tabular}[c]{@{}l@{}}13.40\% ±1.33\%\\ 0.165 ±0.031\end{tabular} & \begin{tabular}[c]{@{}l@{}}11.10\% ±1.24\%\\ 0.068 ±0.019\end{tabular} \\ \hline

\textbf{IPD (All Devices)} & \begin{tabular}[c]{@{}l@{}}Within 0.5mm\\ R²\end{tabular} & \begin{tabular}[c]{@{}l@{}}87\%\\ 0.857\end{tabular} & \begin{tabular}[c]{@{}l@{}}19.47\% ±1.81\%  \\ 0.318 ±0.038\end{tabular} & \begin{tabular}[c]{@{}l@{}}14.17\% ±1.35\%  \\ 0.134 ±0.027\end{tabular} & \begin{tabular}[c]{@{}l@{}}12.17\% ±1.26\%  \\ 0.068 ±0.017\end{tabular} \\ \hline

\textbf{Wingspan} & \begin{tabular}[c]{@{}l@{}}Within 7cm\\ Within 12cm\\ R²\end{tabular} & \begin{tabular}[c]{@{}l@{}}87\%\\ 100\%\\ 0.669\end{tabular} & \begin{tabular}[c]{@{}l@{}}53.93\% ±3.61\%  \\ 78.80\% ±2.76\%\\ 0.134 ±0.042\end{tabular} & \begin{tabular}[c]{@{}l@{}}42.13\% ±3.32\%  \\ 66.00\% ±3.31\%\\ 0.047 ±0.019\end{tabular} & \begin{tabular}[c]{@{}l@{}}40.80\% ±2.80\%  \\ 65.46\% ±3.14\%\\ 0.036 ±0.021\end{tabular} \\ \hline

\textbf{Room Size} & \begin{tabular}[c]{@{}l@{}}Within 2m²\\ Within 3m²\\ R²\end{tabular} & \begin{tabular}[c]{@{}l@{}}78\%\\ 97\%\\ 0.974\end{tabular} & \begin{tabular}[c]{@{}l@{}}22.11\% ±2.85\%\\ 33.52\% ±3.80\%\\ 0.406 ±0.153\end{tabular} & \begin{tabular}[c]{@{}l@{}}16.33\% ±2.74\%\\ 23.44\% ±3.08\%\\ 0.495 ±0.171\end{tabular} & \begin{tabular}[c]{@{}l@{}}12.66\% ±2.98\%\\ 19.53\% ±2.92\%\\ 0.360 ±0.136\end{tabular} \\ \hline

\textbf{Longer Arm} & \begin{tabular}[c]{@{}l@{}}$\geq$ 1cm Difference\\ $\geq$ 3cm Difference\end{tabular} & \begin{tabular}[c]{@{}l@{}}63\%\\ 100\%\end{tabular} & \begin{tabular}[c]{@{}l@{}}58.63\% ±5.79\% \\ 77.78\% ±13.46\%\end{tabular} & \begin{tabular}[c]{@{}l@{}}52.35\% ±6.83\%\\ 62.22\% ±15.09\%\end{tabular} & \begin{tabular}[c]{@{}l@{}}54.90\% ±5.12\%\\ 53.33\% ±15.64\%\end{tabular} \\ \hline

\textbf{Handedness} & Categorical & 97\% & 92.5\% & 75\% & 57.5\% \\ \hline


\textbf{Geolocation} & \begin{tabular}[c]{@{}l@{}}Within 400km\\ Within 500km\end{tabular} & \begin{tabular}[c]{@{}l@{}}50\%\\ 90\%\end{tabular} & \begin{tabular}[c]{@{}l@{}}0\%\\ 6.66\%\end{tabular} & \begin{tabular}[c]{@{}l@{}}0\%\\ 0\%\end{tabular} & \begin{tabular}[c]{@{}l@{}}0\%\\ 0\%\end{tabular} \\ \hline

\textbf{Reaction Time} & Categorical & 87.50\% & 79.20\% & 62.50\% & 54.20\% \\ \hline

\textbf{HMD Refresh Rate} & Within 3 Hz & 100\% & 0\% & 0\% & 0\% \\ \hline

\textbf{Tracking Refresh Rate} & Within 2.5 Hz & 100\% & 0\% & 0\% & 0\% \\ \hline

\textbf{VR Device} & Categorical & 100\% & 10\% & 0\% & 0\% \\ \hline

\multicolumn{6}{l}{} \\
\multicolumn{6}{c}{\textit{Table \ref{tab:results}B: Inferred Attributes (MetaData \cite{nair2022exploring} Study)}} \\
\hline

\textbf{Attribute} & \textbf{Metric} & \textbf{No Privacy} & \textbf{Low Privacy} & \textbf{Medium Privacy} & \textbf{High Privacy} \\ \hline 
\hline

\textbf{Voice} & \begin{tabular}[c]{@{}l@{}}Gender\\ Ethnicity\end{tabular} & \begin{tabular}[c]{@{}l@{}}97\%\\ 63\%\end{tabular} & \begin{tabular}[c]{@{}l@{}}72.5\% ±15\%\\ 52.5\% ±7.5\%\end{tabular} & \begin{tabular}[c]{@{}l@{}}65\% ±15\%\\ 40\% ±5\%\end{tabular} & \begin{tabular}[c]{@{}l@{}}61.25\% ±13.75\%\\ 32.5\% ±0.5\%\end{tabular} \\
\hline

\textbf{Gender} & Categorical & 100\% & 76.5\% ±1.29\% & 70.47\% ±1.85\% & 57.19\% ±2.20\% \\
\hline

\textbf{Age} & Within 1yr & 100\% & 41.75\% ±1.65\% & 36.09\% ±1.87\% & 24.28\% ±1.87\% \\
\hline

\textbf{Ethnicity} & Categorical & 100\% & 51.25\% ±2.70\% & 40.75\% ±2.36\% & 31.37\% ±2.40\% \\
\hline

\textbf{Income} & Within \$10k & 100\% & 26.15\% ±1.41\% & 28.00\% ±1.87\% & 26.06\% ±2.11\% \\
\hline

\multicolumn{6}{l}{} \\
\multicolumn{6}{c}{\textit{Table \ref{tab:results}C: Identity (TTI \cite{millerpersonal2020}, MetaData \cite{nair2022exploring}, and 50k \cite{nair2023unique} Studies)}} \\
\hline

\textbf{Attribute} & \textbf{Dataset} & \textbf{No Privacy} & \textbf{Low Privacy} & \textbf{Medium Privacy} & \textbf{High Privacy} \\ \hline 
\hline

\textbf{Identity} & TTI (Miller et al.) & 95\% & 81.10\% ±5.78\% & 45.29\% ±5.48\% & 26.51\% ±1.37\% \\
\hline

\textbf{Identity} & MetaData (Nair et al.) & 100\% & 5.44\% ±0.68\% & 4.59\% ±0.76\% & 4.0\% ±0.67\% \\
\hline

\textbf{Identity} & 50k (Nair et al.) & 94.33\% & 15.59\% ±4.50\% & 6.10\% ±1.76\% & 2.19\% ±1.17\% \\
\hline

\end{tabular}%
}
\vspace{1em}
\caption{Main Results (accuracy and R² values with $\mathbf{99\%}$ confidence intervals)}
\label{tab:results}
\end{table}

\end{document}